\documentclass[referee]{aa}

\usepackage{lscape}
\usepackage{graphicx}
\usepackage{txfonts}
\usepackage{longtable}
\usepackage{natbib}

\begin{document}

\title{The Hawaii Trails Project: Comet-Hunting in the Main Asteroid Belt
       \thanks{
       Some of the data presented herein were obtained at the W. M.
       Keck Observatory, the Gemini Observatory, Subaru Telescope,
       National Optical Astronomy Observatory (NOAO) facilities at
       the Cerro Tololo Inter-American Observatory, and Lulin Observatory.
       Keck is operated as a scientific partnership among the California
       Institute of Technology, the University of California, and the
       National Aeronautics and Space Administration, and was made possible
       by the generous financial support of the W. M. Keck Foundation.
       Gemini is operated by the Association of Universities for Research
       in Astronomy, Inc., under a cooperative agreement with the National
       Science Foundation (NSF) on behalf of the Gemini partnership.
       Subaru is operated by the National Astronomical Observatory of Japan.
       NOAO and Cerro Tololo are operated by the Association of Universities for
       Research in Astronomy, Inc., under co-operative agreement with the NSF.
       Lulin is supported and was made possible
       by the National Science Council of Taiwan, the Ministry of Education
       of Taiwan, and National Central University.}
}

\author{Henry H. Hsieh\inst{1,2}}

\institute{
    Astrophysics Research Centre, Queen's University,
    Belfast, BT7 1NN, United Kingdom\\
    \email{h.hsieh@qub.ac.uk} \and
    Institute for Astronomy, University of Hawaii, 2680 Woodlawn
    Drive, Honolulu, HI 96822, USA
}


\date{Submitted 2009 April 16; Accepted 2009 July 30}

\begin{abstract}
{The mysterious solar system object 133P/(7968) Elst-Pizarro is
 dynamically asteroidal, yet displays recurrent comet-like dust emission.
 Two scenarios were hypothesized to explain this unusual
 behavior: (1) 133P is a classical comet from the outer
 solar system that has evolved onto a main-belt orbit, or
 (2) 133P is a dynamically ordinary main-belt asteroid on
 which subsurface ice has recently been exposed.  If (1) is correct,
 the expected rarity of a dynamical transition
 onto an asteroidal orbit implies that 133P could be alone in the
 main belt.  In contrast, if (2) is correct,
 other icy main-belt objects should exist and could
 also exhibit cometary activity.}
{Believing 133P to be a dynamically ordinary, yet icy main-belt
 asteroid, I set out to test the primary prediction of the hypothesis:
 that 133P-like objects should be common and could be found
 by an appropriately designed observational survey.}
{I conducted just such a survey --- the Hawaii Trails Project
 --- of selected main-belt asteroids in a search for
 objects displaying cometary activity.  Optical observations were
 made of targets selected from among
 the Themis, Koronis, and Veritas asteroid families, the Karin asteroid cluster,
 and low-inclination, kilometer-scale
 outer-belt asteroids, using the Lulin 1.0~m,
 Small and Moderate Aperture Research Telescope System (SMARTS) 1.0~m,
 University of Hawaii 2.2~m, Southern Astrophysical Research (SOAR) 4.1~m,
 Gemini North 8.1~m, Subaru 8.2~m, and Keck I 10~m telescopes.
}
{I made 657 observations of 599 asteroids, discovering one
 active object now known as 176P/LINEAR, leading
 to the identification of the new cometary class of main-belt comets.
 These results suggest that there could be $\sim$100 currently active
 main-belt comets among low-inclination, kilometer-scale outer belt asteroids.
 Physically and statistically, main-belt comet activity
 is consistent with initiation by meter-sized impactors.  The
 estimated rate of impacts and sizes of resulting active sites, however, imply
 that 133P-sized bodies should become significantly devolatilized over Gyr
 timescales,
 suggesting that 133P, and possibly the other MBCs as well,
 could be secondary, or even multigenerational, fragments from recent breakup
 events.
}
{}
\end{abstract}


\titlerunning{The Hawaii Trails Project}
\authorrunning{H. H. Hsieh}

\maketitle

\section{Introduction}

\subsection{Motivation\label{motivation}}

Comet 133P/(7968) Elst-Pizarro (hereafter 133P), an apparent Themis asteroid
family member, has an asteroid-like Tisserand invariant of $T_J=3.16$ and
orbits in the
outer main asteroid belt. Its cometary nature was first revealed on 1996
August 7 when a linear dust feature was observed trailing the object
\citep{els96}.  This dust trail was found to be narrow, structureless, and
over 3 arcmin in length.  No coma was observed around the nucleus.  Previous
images from 1979 and 1985 \citep{mar96,mcn96} and follow-up observations by
\citet{off97} revealed a completely point-source-like nucleus with no
apparent dust trail.

The uniqueness of 133P's dust emission --- no other asteroid had
ever been seen exhibiting such behavior, and even 133P only
demonstrated such behavior once --- made it difficult to ascertain
its true nature.  Some believed the emission to be impact
ejecta from a collision between 133P and another asteroid or group of
asteroids \citep[{\it e.g.},][]{tot00}, while
others believed it to be the result of the sublimation of volatile
ices on 133P's surface \citep[{\it e.g.},][]{boe98}.
Impact scenarios were unable to plausibly explain the months-long duration of
the dust emission episode \citep{boe96}, however, while sublimation
required the apparently implausible scenario of surface ice
surviving to the present-day on a main-belt asteroid in quantities sufficient
for driving cometary activity.

New insight was gained into 133P's nature when observations
in 2002 revealed the return of long-lived dust emission
\citep{hsi04}.  Given the implausibility of impacts
causing prolonged dust emission episodes on the same asteroid in
the span of six years,
these observations effectively ruled out hypotheses in which
impacts were the sole cause of the observed dust emission.
After considering other possible
emission mechanisms and ruling out each in turn,
Hsieh {\it et al.} concluded that 133P's dust ejection was most
likely driven by the sublimation of volatile material, presumed
to be water ice.

This conclusion implied either that (1) 133P was a ``lost comet'', {\it i.e.},
an object originally
from the outer solar system that had evolved onto its current orbit
via planetary encounters, dynamical resonances, or the
non-gravitational influence of asymmetric cometary mass loss, or (2)
133P was an ``icy asteroid'', {\it i.e.}, a dynamically ordinary, native
member of the asteroid belt on which preserved, buried ice had been
recently exposed, perhaps excavated by an impact.
If the ``lost comet'' hypothesis was correct,
the low likelihood of a comet undergoing a dynamical
transition from the outer solar system onto a 133P-like orbit
\citep[{\it e.g.},][]{ipa97,fer02} could mean that such
objects are rare or even non-existent, and that 133P could
be the only object of its kind, as was observed. If the
``icy asteroid'' hypothesis was correct, 133P would not be expected to be unique.
Other main-belt asteroids could also possess subsurface ice
that could begin sublimating if exposed by impacts.
In this case, 133P's uniqueness could simply be due to
the difficulty of discovering such objects serendipitously.
An appropriately targeted search could counter this difficulty \citep{hsi06a}.
Believing that this second hypothesis could be correct, I set out to conduct
just such a targeted search.  In this paper, I describe the design and results of
that survey, the Hawaii Trails Project (HTP).

\subsection{Background\label{whynoaas}}

Various evidence has been found for past and even present water in main-belt
asteroids \citep[{\it cf}.][and references within]{hsi06a}, but this fact has not
meant that active dust emission due to the sublimation of
water ice is commonly observed.  Hsieh \& Jewitt discuss a number of
possible reasons for this inconsistency.
Perhaps simply too few 133P-like objects had been observed
sufficiently deeply and sufficiently often.
Collisional excavation of subsurface ice reservoirs could
occur only rarely.  Even when collisions do occur,
sublimation may not necessarily follow due to the probable
non-uniform distribution of subsurface ice reservoirs, and even the successful triggering
of sublimation could result in dust emission that is simply too weak to detect
from Earth.
Finally, observable dust emission is likely to be intermittent and also have a
finite lifetime, limited by either local devolatilization of the active site
or mantling \citep[][and references within]{jew96,jew02}, further reducing
the probability of its detection.

In addition to the above considerations, there are indications that
detectable activated asteroids may occupy a narrow range of sizes.
The rate of impact excavations of an asteroid depends on
its collisional cross-section:
large bodies will be struck, and thus activated, more often than small bodies.
Volatile material should also survive longer against solar heating
on larger asteroids because it can be
buried at greater depths.
However, \citet{hsi04} found that the low dust ejection
velocity ($v_{ej}\sim 1-2$~m~s$^{-1}$) for 133P was extremely similar to
the gravitational escape velocity ($v_{esc}\sim 1.5$~m~s$^{-1}$),
suggesting comparable dust emission would only be observable on bodies
of similar or smaller sizes.  Comparable emission on larger bodies would
be unable to escape the gravity of those bodies and thus never become observable.
This scenario may explain why Ceres is not observed to emit dust,
despite indications of surface ice \citep{leb81,ver05} and
even possible water vapor emission \citep{ahe92}.
Despite providing less insulation from the Sun than larger bodies,
smaller bodies are also less effectively heated from within by $^{26}$Al
(since they radiate that heat away more efficiently).
In summary, to exhibit dust emission, it seems that
icy asteroids need to be large enough to preserve subsurface ice
against solar heating and be struck relatively frequently by impactors, but
not so large that weak dust emission cannot escape gravity or ice does not
survive early radioactive heating.
Such conflicts in size preferences suggests that
133P-like asteroids may be even rarer and more difficult to discover than
initial considerations might indicate.

\subsection{Survey Design\label{surveydesign}}

The distribution of present-day ice in the asteroid belt is not well understood.
The so-called snow line refers to the distance from the Sun at which the
temperature of the protosolar disk was below the condensation temperature
of water, allowing ice to become incorporated into accreting planetesimals.
Observations of asteroids suggest that the snow line likely existed
$\sim$2.5~AU from the Sun \citep{gra82,jon90}, though theoretical
models have suggested it may have been as close as the orbit of Mars
\citep{sas00,cie06,lec06}, suggesting objects throughout the asteroid
belt could have formed icy.  Much of this ice has since undergone thermal
processing, producing the hydrated minerals now observed on many main-belt
asteroids \citep[see][]{riv02}.  Detections of such minerals decline in
the outer belt beyond $\sim$3~AU (notably, where 133P is found),
suggesting that pristine, unaltered ice could still exist in
distant main-belt objects \citep{jon90,sco05}.  This hypothesis suggests that
a search for 133P-like icy asteroids would be best served by focusing on
objects with semimajor axes of $a>3$~AU, as the HTP does indeed do,
although objects with $a<3$~AU are also considered.

Given the environmental and physical conditions discussed above that are
suspected of giving rise to cometary activity in 133P, in the search for objects
exhibiting similar activity, the following target categories were considered:

\begin{itemize}
\item{{\bf Themis family asteroids:} The Themis family is likely
  the result of the breakup of a parent asteroid about 400~km
  in diameter $\sim$2.5~Gyr ago \citep{mar95,tan99,nes03} and is one of
  the largest and most statistically robust asteroid families known
  \citep[{\it e.g.},][]{car82,zap90}. Given their origin from a
  common parent, Themis family members are thought to be
  compositionally homogeneous, as corroborated
  by observational studies showing that the family is dominated by
  primitive C-type asteroids that
  also exhibit signs of aqueous alteration
  \citep{bel89,flo99,ive02,mot05}.  The family is also characterized by a
  high rate of collisions relative to the general main-belt population
  \citep{far92a,del01b}.
  Given 133P's dynamical and possible compositional similarity to Themis
  family members and the high frequency of collisions within the family,
  the Themis family was judged to be a natural place to look
  for 133P-like asteroids. Target objects were selected by
  conducting a hierarchical clustering analysis \citep{zap90,zap94} on
  proper elements obtained from the AstDys website
  \citep[http://hamilton.dm.unipi.it/astdys/ ;][]{kne03b} and using a
  cutoff value of $\delta v'=70$~m~s$^{-1}$ with respect to (24) Themis.
  }
\item{{\bf Low-inclination outer-belt asteroids:} The identification of
  the main-belt comet (MBC) class \citep{hsi06b} during the course of the HTP
  led to a re-evaluation of the survey's target selection criteria.
  In particular, the semimajor axis and
  inclination of MBC P/2005 U1 (Read) (hereafter P/Read)
  closely matched those of the Themis
  family, but its eccentricity did not.  This discovery led to the hypothesis
  that 133P's low-inclination, and not necessarily its specific association
  with the Themis family, could be primarily responsible for its activity.
  Collision rates are in fact known
  to be enhanced at low inclinations \citep{far92a}.  Targets were selected by
  applying limits (3.0~AU $<a<3.3$~AU, $e<0.3$, and $i<3.0^{\circ}$) on
  osculating orbital elements obtained online from Lowell Observatory
  (ftp://ftp.lowell.edu/pub/elgb/astorb.dat),
  permitting the inclusion of even recently discovered asteroids for which
  proper elements were not yet known.  Limits were chosen to span the orbital
  element distribution of the first three
  known MBCs, and in fact, also completely encompasses the Themis family.
  }
\item{{\bf Koronis family asteroids:} As another populous main-belt asteroid
  family characterized by high collisions rates
  \citep{far92a}, the Koronis family was judged to
  be another plausible place to look for 133P analogs.  Like the Themis
  family, the Koronis family is the result of the catastrophic disruption
  of a large parent body $\sim$2~Gyr ago \citep{bin88,mar95}.
  Unlike the Themis family, however, the Koronis family is dominated by
  S-type asteroids \citep{bel89,bin93,mot05}, which along with their
  meteoritic analogs, the ordinary chondrites, do not exhibit
  spectroscopic evidence of significant aqueous alteration
  \citep[{\it e.g.},][]{riv02}.  
  Nonetheless, due to the high collision rates within this
  family and the possibility that these asteroids may not actually be
  completely anhydrous \citep{gro00,kei00}, Koronis
  family asteroids were still judged to be viable 133P-analog candidates. Koronis family
  targets were selected by a hierarchical clustering analysis, using a cutoff value
  of $\delta v'=50$~m~s$^{-1}$ with respect to (158) Koronis.
  }
\item{{\bf Karin cluster asteroids:} The Karin cluster is contained
  within the Koronis family and is likely
  the result of a recent fragmentation event $\sim$5.8~Myr ago
  \citep{nes02,nes06}.
  Spectroscopic studies have since affirmed both the compositional
  homogeneity and the young age of the cluster
  \citep{sas04,bru06,ver06}.
  In the context of this survey, the Karin cluster was interesting because
  if its parent body only recently broke apart, ice that had been preserved
  deep inside could suddenly be
  located at a much shallower depths on the resulting fragments.
  Even relatively weak subsequent impacts could then easily initiate sublimation.
  Like the rest of the Koronis family, however, the Karin cluster is dominated
  by S-type asteroids and so is not an ideal sample from a mineralogical point
  of view for searching for sublimation-driven dust emission.
  Given the likelihood of particularly young surfaces, though, these objects
  were judged to be worthy of at least
  cursory attention.  Karin cluster targets were
  selected by a hierarchical clustering analysis, using
  a cutoff value of $\delta v'=10$~m~s$^{-1}$ with respect to (832) Karin.
  }
\item{{\bf Veritas family asteroids:} Like the Karin cluster, the
  Veritas family, first identified
  by \citet{zap95}, is likely the result of a recent breakup
  in the asteroid belt, perhaps just $\sim$8.5~Myr ago \citep{nes03,tsi07}.
  Intriguingly, such a breakup corresponds
  with a spike in $^3$He concentrations found in sea-floor sediment
  records from that time period \citep{far06}.
  The young age of the Veritas family made it interesting for this survey for
  the same reasons as the Karin cluster.
  Unlike the Karin cluster, however, the Veritas family appears to be
  dominated by C-type asteroids \citep{mot05}, the same type of
  asteroids that dominate the Themis family, making it a particularly promising
  place to search for 133P analogs.  Veritas family targets were
  selected by a hierarchical clustering analysis, using a cutoff
  value of $\delta v'=40$~m~s$^{-1}$ with respect to (490) Veritas.
  }
\end{itemize}

Targets from these categories were selected for each observing run based
on visibility, expected visual magnitude ($18\la m_R \la20$~mag
targets for UH 2.2~m runs, brighter targets for smaller telescopes or
in non-photometric conditions, and fainter targets for larger telescopes),
distance from the moon, and distance from the galactic plane.
I also sought to observe asteroids matching 133P closely in physical size
(as parameterized by absolute magnitude, $H_V$), or approximately
kilometer-scale, given the issues discussed in \S\ref{whynoaas}.
The aim was to achieve moderately deep imaging for as many asteroids as possible.
As such, exposure times were selected such that signal-to-noise
ratios of at least $S/N=20$ for 93.5\% (614 out of 657) of the surveyed targets,
and at least $S/N=10$ for 99.2\% (652 out of 657) of the surveyed targets were reached.

While most observations were made of unique asteroids, repeat observations
of individual asteroids were not expressly avoided.  Given the expected transient
nature of any cometary activity, continued monitoring of comet candidates
is in fact, in principle, vital to fully characterizing their active natures.
In a larger survey, all objects would have been observed multiple times, but for the
relatively small-scale HTP, repeat observations were made only occasionally.

\section{Observations\label{observations}}

Several ground-based optical observatories were used
to conduct this survey: the University of Hawaii (UH) 2.2~meter
telescope, the 8.1~m Gemini North Observatory, the 8.2~m Subaru telescope,
and the 10~m Keck I Observatory on Mauna Kea in Hawaii, the
1.0~m telescope operated by the Small and
Moderate Aperture Research Telescope System (SMARTS) consortium
at the Cerro Tololo Inter-American Observatory (CTIO) and Cerro Pachon's 4.1~m
Southern Astrophysical Research (SOAR) telescope in
Chile, and the Lulin 1.0~m telescope in Taiwan
(Table~\ref{telescopes}).
Data was obtained on 88 separate nights (Table~\ref{obsruns}).

Either a Tektronix 2048$\times$2048 pixel charge-coupled device (CCD)
or the Orthogonal Parallel Transfer Imaging Camera (OPTIC) was used at the f/10 focus
of the UH 2.2 m telescope.
Other instruments used on Mauna Kea telescopes were the Gemini Multi-Object
Spectrograph (GMOS) \citep{hoo04} in imaging mode on Gemini,
the Subaru Prime Focus Camera (Suprime-Cam) \citep{miy02} on Subaru, and
the Low-Resolution Imaging Spectrometer (LRIS) \citep{oke95} on Keck.
Observations using the CTIO 1.0~m telescope employed either an
Apogee 512x512 CCD or an STA 4064x4064 CCD (Y4KCam), while observations on
the SOAR telescope employed the SOAR Optical Imager (SOI) \citep{sch04}.
Observations using the Lulin 1.0~m telescope were made using a VersArray:1300B
CCD \citep{kin05}.
All observations were obtained through broadband filters approximating the
Kron-Cousins $BVRI$ photometric system except for observations at Gemini
which employed $g'r'i'z'$ broadband filters similar to those used for the
Sloan Digital Sky Survey (SDSS).
Further telescope details are listed in Table~\ref{telescopes}.

Standard image preparation (bias subtraction and flat-field
reduction) was performed on all data.  Flat fields were constructed either
from dithered images taken nightly of the twilight sky or images of the
illuminated interior of the telescope dome.  Photometry on
standard stars and target objects was performed by measuring net fluxes
contained within circular apertures with optimum sizes determined from
curve of growth analysis and dependent on the nightly seeing.  The sky
background to be subtracted was determined from the median pixel value
within a circular annulus surrounding each central aperture.
Photometry was calibrated to \citet{lan92} standard stars to produce absolute
photometry for the target objects.
Photometric uncertainties are estimated to be 0.1~mag for objects
observed in photometric conditions and 0.5~mag for objects observed
in non-photometric conditions (see Table~\ref{obsruns}).

\section{Analysis and Results\label{analysisresults}}

\subsection{Activity Assessment\label{cometassessment}}

To search for low-level cometary activity, all images for a given target
on a single night were aligned on the object's photocenter using linear
interpolation and then summed to produce a single composite image.
Asteroids with possible cometary activity were identified visually from
these composite images, and then the corresponding individual images were
inspected to assess the plausibility of the detection ({\it e.g.}, whether
the morphology of a suspected cometary feature matched that expected
for comets, whether the feature consistently appeared in each individual
image, whether the motion of the feature and object against the background
star field were identical, and
whether there were any signs that the features could be attributed to
scattered light or imperfect flatfielding). After this close inspection,
if an object showed credible evidence of being active, confirmation of
activity was then attempted using deep follow-up imaging.
Observations for which activity was impossible to assess and were therefore
unsuitable for analysis ({\it e.g.} those where the object was strongly affected
by scattered light from
a nearby bright star, or where poor 
telescope stability resulted in inconsistent point-spread functions among
individual images of the same object) were discarded.

Various systematic, quantitative means of searching for cometary activity
in HTP data were considered, including one-dimensional surface brightness
profile analysis \cite[{\it e.g.},][]{luu92,hsi05},
radial surface brightness profile analysis \citep[{\it e.g.},][]{luu90},
and trail searching \citep{hsi05}.
One-dimensional surface brightness profile analysis can be useful
in that it accounts for trailing of point source templates ({\it i.e.}, field
stars) caused by non-sidereal tracking and long exposure
times of each target image.
Trailing effects are avoided by measuring the surface brightness
profiles of targets and field stars in the direction perpendicular to the
apparent motion of the target.  This technique is best-suited for detecting
coma that extends radially in all directions from an object, or directed
emission not aligned with the direction of the object's apparent
motion, but would miss directed emission oriented along the direction of
an object's apparent motion (as was observed for 133P).

Radial surface brightness profile analysis is sensitive to cometary
features extended in any direction, but requires the use of circularly-symmetric
point source templates.  Field stars from target
images were generally unsuitable due to trailing, requiring the use of
either standard star images or separately-observed sidereally-tracked
target images to obtain reference star templates, but both of these alternatives
were problematic as well.  Even with
the use of tip-tilt guiding, continuous atmospheric fluctuations
and small but non-negligible mechanical
instabilities in the telescopes used (particularly while tracking
non-sidereally) resulted in systematically wider point-spread functions (PSFs)
for long-exposure target images than 
short-exposure standard star images.  Those same image quality fluctuations
also adversely affected the utility of separate sidereally-tracked target images
by making it difficult to distinguish ordinary image-to-image PSF fluctuations from
the small deviations that could actually indicate the presence of activity.
This technique also averages any dust emission
over all radial directions, reducing sensitivity to directed emission.

Searching for dust emission specifically like 133P's dust trail, which could form if
an object emits large ($\ga10$~$\mu$m) dust particles at low velocities
($\sim1-2$~m~s$^{-1}$) as 133P did \citep{hsi04}, can be done by searching
for excess flux above sky background along the object's orbital plane
(the expected direction of a trail).
This technique is not sensitive to
near-nucleus activity, however, and furthermore, is highly susceptible to
contamination from field stars or galaxies due to the large amount of
sky that must be examined for evidence of trails.

All three techniques are prone to false detections, which
can arise from nearby faint field stars or galaxies masquerading as active features,
contamination of sky background samples, imperfect
flatfielding, scattered light from nearby bright or saturated field stars, and
PSF variability between images or even within the same image ({\it i.e.},
as a function of position on the CCD).
Thus, any automated detection scheme still requires confirmation by visual
inspection.  Given the focused nature of this survey (targeted deep imaging of
single objects), the resulting small sample size ($\sim$650
composite images to be evaluated over 3 years), and the
significant human intervention needed to supplement any of the quantitative
detection algorithms considered, I opted to simply visually
inspect each target for evidence of activity.  This approach
is certainly less feasible for searching for comets in much larger data
sets, particularly the ones that will be produced by upcoming survey telescopes
like the Panoramic Survey Telescope and Rapid Response System
\citep[Pan-STARRS;][]{hod04} and the Large Synoptic Survey Telescope \citep[LSST;][]{jon09}.
For the HTP, however, direct visual inspection was judged to be the most
efficient (given the small sample size) and effective
method for evaluating the active nature of the targets in this survey.

\subsection{Survey Results\label{surveyresults}}

In total, 599 unique asteroids were observed as part of the HTP,
with 544 asteroids observed once, 52 asteroids observed twice, and
3 asteroids observed three times, totaling 657 separate observations
(Table~\ref{objprops}).  Hereafter, all observations (including those
of objects observed multiple times) are considered independent of one
another.  This approach is taken since it is not known
where along their orbits other 133P-like asteroids could exhibit
activity, and also since inactivity observed for an asteroid
at one time does not necessarily rule out activity at another
time.

\begin{figure}
\includegraphics[width=\textwidth]{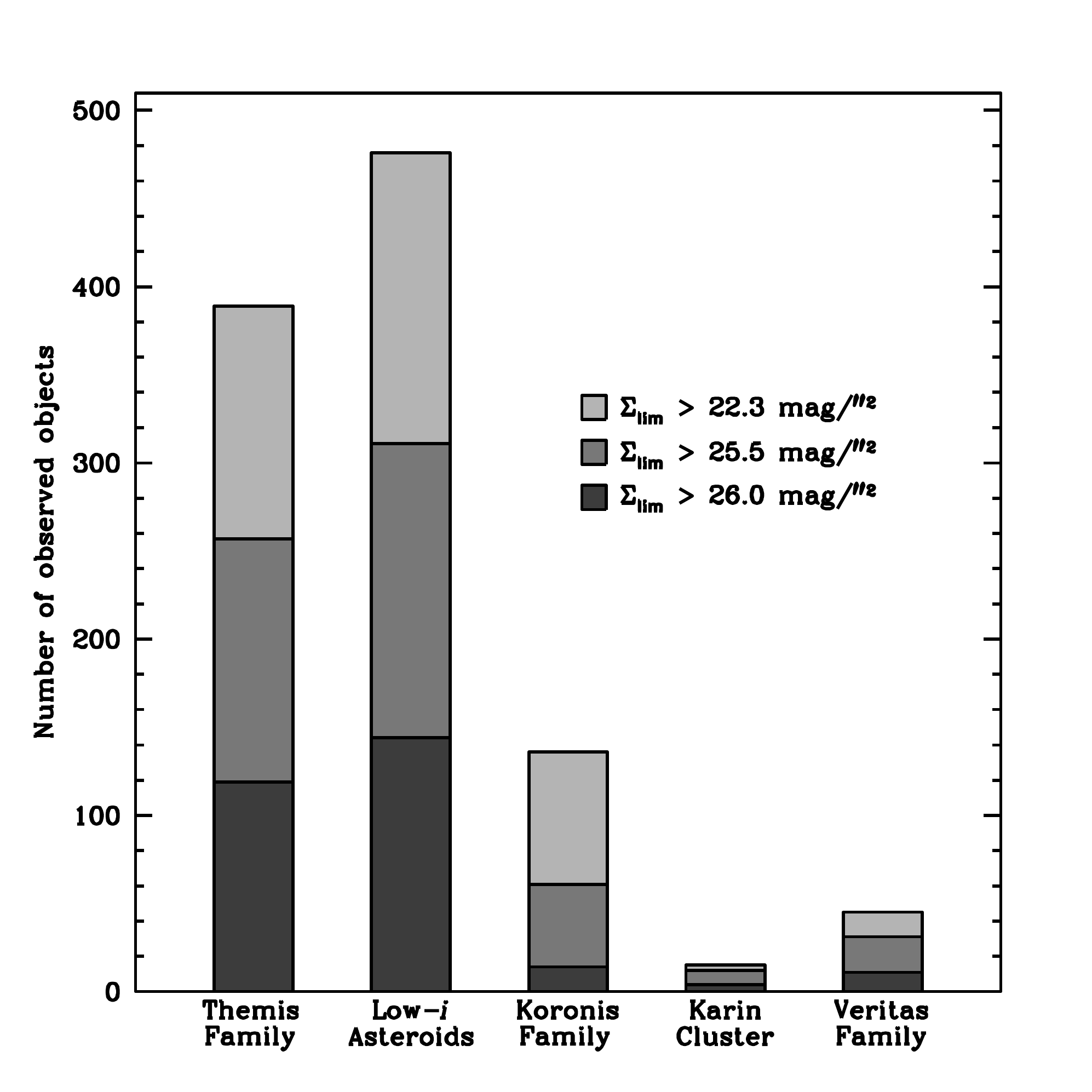}
\caption{Distribution of observed objects by target category, 
  where all Themis family members are also low-inclination outer-belt asteroids,
  and all Karin cluster members are also Koronis family members.
  Observations for which activity detection limits are fainter than 22.3~mag~arcsec$^{-1}$,
  25.5~mag~arcsec$^{-1}$, and 26.0~mag~arcsec$^{-1}$
  are shown in light gray, medium gray, and dark gray, respectively.
  }
\label{survey_types}
\end{figure}

The numbers of targets observed from each target category are shown in
Table~\ref{surveysummary} and Fig.~\ref{survey_types}, for which it
should be noted that all Themis family members are also counted as
low-inclination outer belt asteroids, and all Karin cluster members are
also counted as Koronis family members.
The Veritas family is independent of all other target categories.
Slightly more than half (389 of 657) of observed targets were
Themis family members, reflecting the prediction that these were the
objects among which other 133P-like asteroids would most likely be found.
This histogram, and others following it, also show the relative numbers of
targets for which shallow, medium, and deep surface
brightness detection limits were reached.

\begin{figure}
\includegraphics[width=\textwidth]{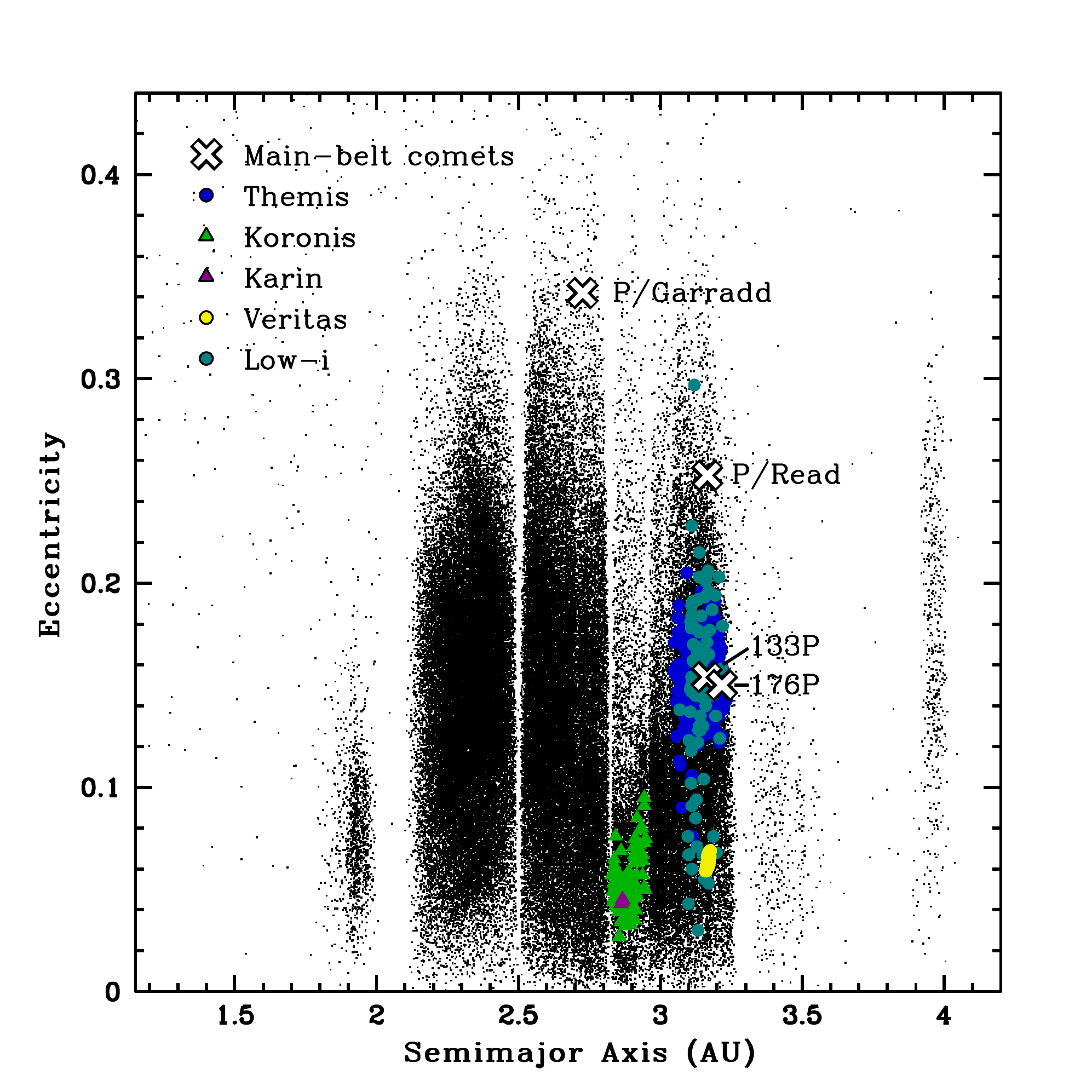}
\caption{Semimajor axis versus eccentricity distribution of
  HTP survey targets, where proper orbital elements are plotted when available and
  osculating orbital elements are plotted for objects for
  which proper orbital elements are not available.  Proper orbital elements for the
  background main-belt asteroid population are plotted using small black dots, while
  orbital elements for observed HTP targets are marked as labeled.
  Orbital elements of the currently-known MBCs are plotted for reference, although
  only 176P was observed as part of this survey.
  }
\label{survey_ae}
\end{figure}

\begin{figure}
\includegraphics[width=\textwidth]{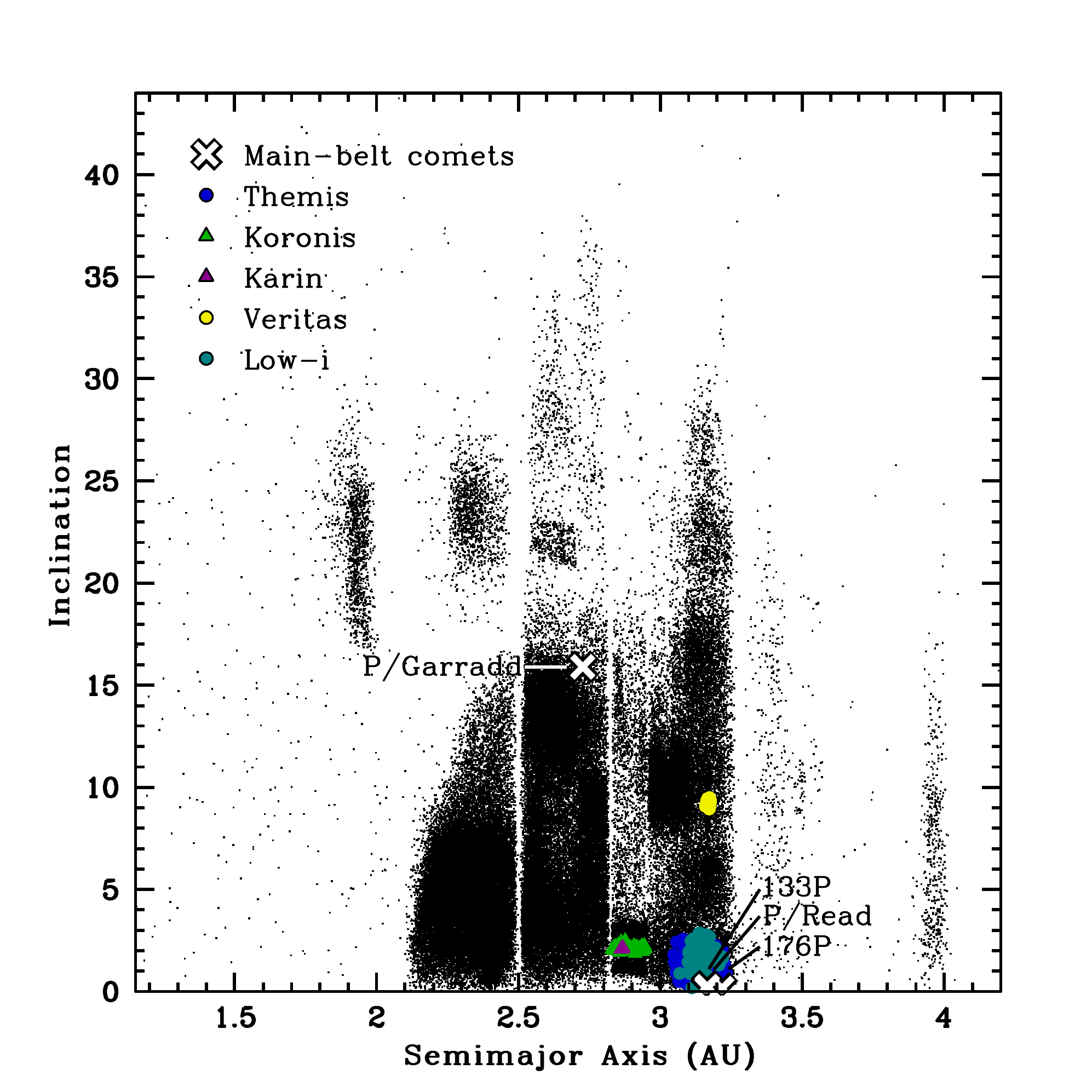}
\caption{Semimajor axis versus inclination distribution of
  HTP survey targets, where proper orbital elements are plotted when available and
  osculating orbital elements are plotted for objects for
  which proper orbital elements are not available.  Proper orbital elements for the
  background main-belt asteroid population are plotted using small black dots, while
  orbital elements for observed HTP targets are marked as labeled.
  Orbital elements of the currently-known MBCs are plotted for reference, although
  only 176P was observed as part of this survey.
  }
\label{survey_ai}
\end{figure}

The semimajor axis versus eccentricity distribution and
semimajor axis versus inclination distribution of surveyed targets in the context
of the background main-belt population are shown in Figs.~\ref{survey_ae} and \ref{survey_ai},
respectively.  These distributions primarily reflect the family association
or orbital element criteria of HTP target categories (\S\ref{surveydesign}).

\subsection{Discovery of the Main-Belt Comets\label{mbcdiscovery}}

The most significant result from the HTP was the
discovery of cometary activity in asteroid 118401 (1999 RE$_{70}$;
subsequently designated 176P/LINEAR; hereafter 176P),
on 2005 November 26 by Gemini North
\citep{hsi06b,hsi06c}.  Coincidentally, this discovery was made
just a month following the serendipitous
discovery of P/Read \citep{rea05},
another cometary object orbiting completely within the main asteroid
belt, on 2005 October 24
by the 0.9~m Spacewatch telescope.  Together, 
133P, P/Read, and 176P formed a new class of objects which were dubbed
MBCs \citep{hsi06b}.
Following the completion of the HTP, a fourth MBC, P/2008 R1 (Garradd)
\citep[hereafter, P/Garradd;][]{gar08}, was discovered on 2008 September 2
by the 0.5-m Uppsala Schmidt telescope at Siding Spring.
The properties of the currently-known MBCs are listed in Table~\ref{mbcprops}.
All other HTP observations
(including a previous observation of 176P using the Lulin 1.0~m telescope
one month prior to the discovery of the object's cometary nature by Gemini)
were judged to show no evidence of cometary activity.

The extremely limited data set used to discover the currently-known
population (three members were discovered serendipitously by 0.5~m to 1.0~m
telescopes, and another was discovered by a targeted survey composed of only 657
asteroid observations) implies that a much larger undiscovered population exists.
The existence of multiple known MBCs and the implication of a much larger
as-yet undiscovered population runs counter to the prediction made by
the ``lost comet'' hypothesis that 133P should be alone in the main belt,
making the ``icy asteroid'' hypothesis, which
predicts that such objects should be
widespread in the main belt, much more likely (\S\ref{motivation}).

\section{Discussion}

\subsection{Selection Effects and Biases\label{seleffects}}

It should be emphasized that the HTP was
not designed as a random sampling of the main asteroid belt and cannot be
interpreted as such.  On the contrary, it was designed to
maximize the probability of discovering 133P-like asteroids and so employed
very specific physical and dynamical target selection criteria.
The HTP was also subject to a number of ordinary observational biases.

\begin{figure}
\includegraphics[height=190pt]{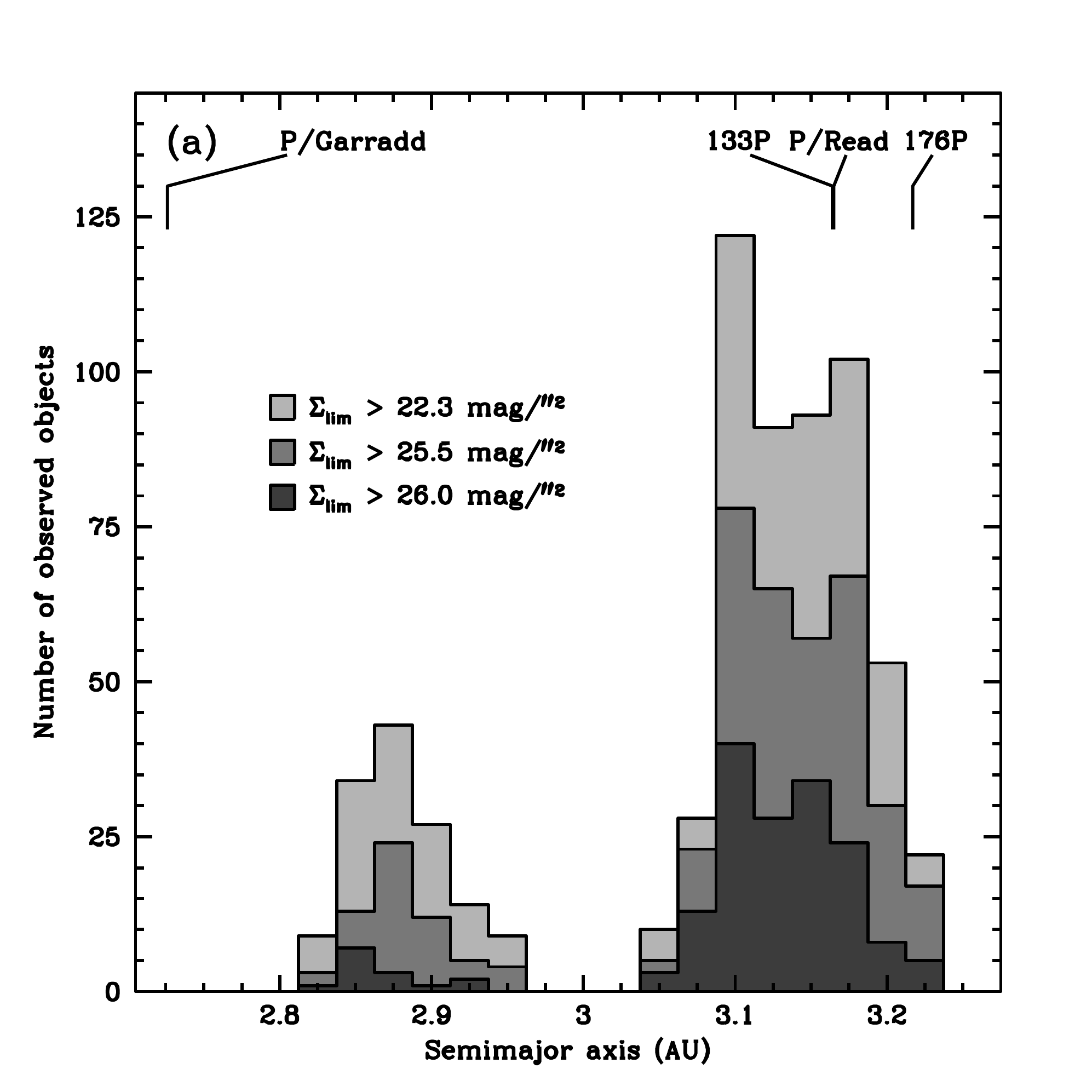}\quad
\includegraphics[height=190pt]{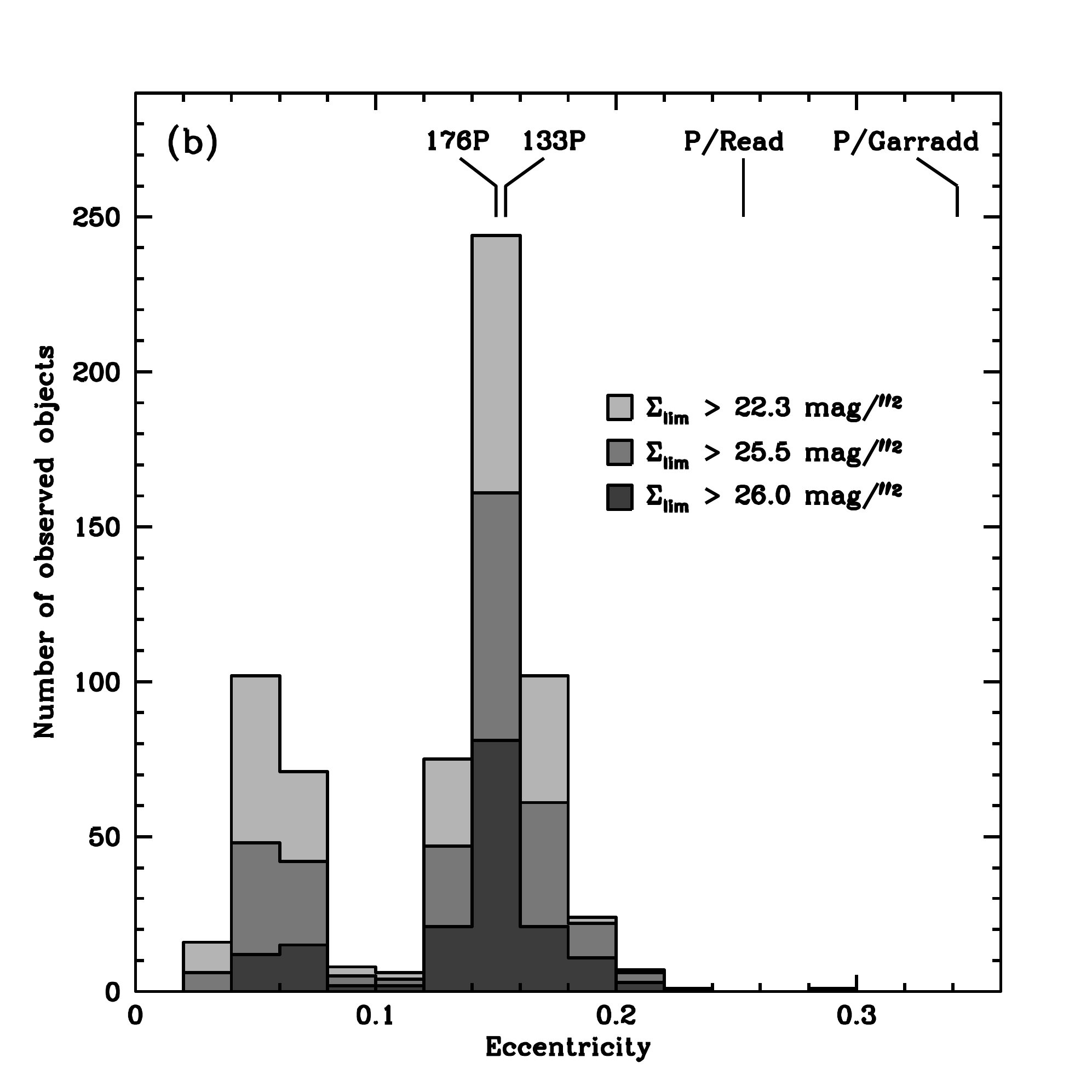}
\includegraphics[height=190pt]{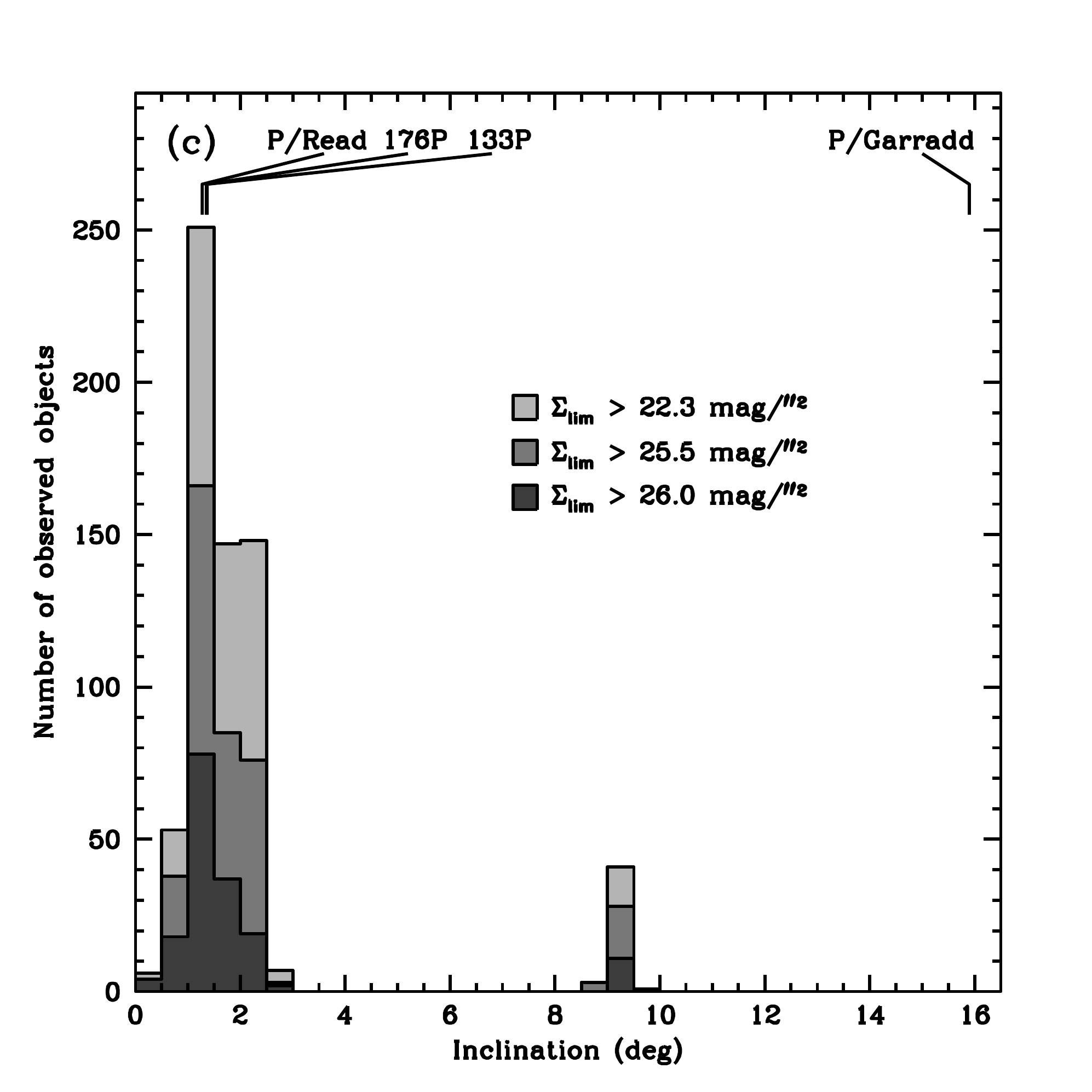}\quad
\includegraphics[height=190pt]{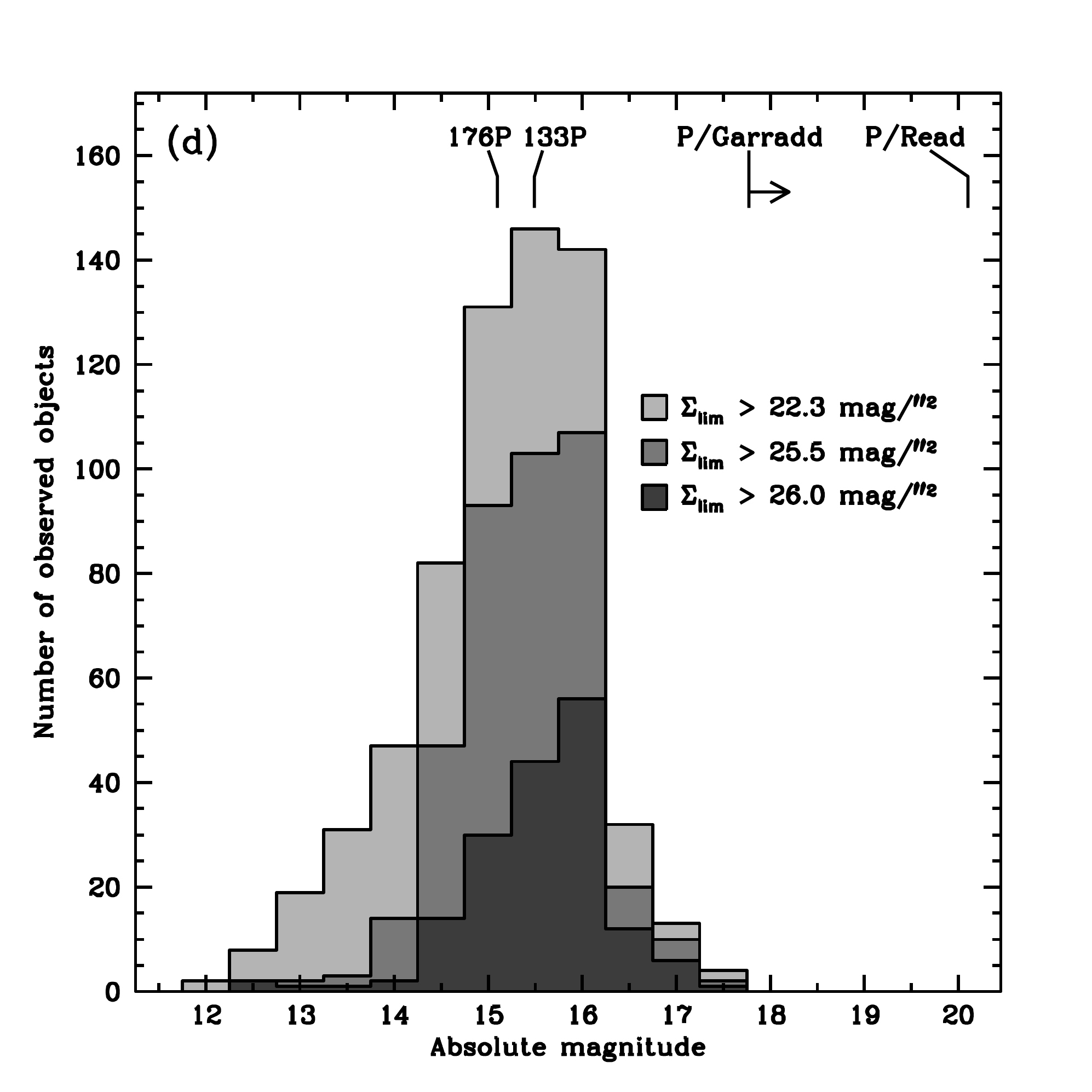}
\caption{Orbital and physical properties of surveyed targets, including
  (a) semimajor axis ($a$) in AU, (b) eccentricity ($e$), (c) inclination ($i$)
  in degrees, and (d) absolute magnitude ($H_V$).  Proper orbital elements are
  used when available, while osculating orbital elements are used for objects
  for which proper elements are not available.  Properties of the currently-known
  MBCs (where the absolute magnitude of P/Garradd is a lower limit)
  are marked for reference, although only 176P was observed as part of this
  survey.
  Observations for which activity detection limits are fainter than 22.3~mag~arcsec$^{-1}$,
  25.5~mag~arcsec$^{-1}$, and 26.0~mag~arcsec$^{-1}$
  are shown in light gray, medium gray, and dark gray, respectively.
  }
\label{orbphysproperties}
\end{figure}

\begin{figure}
\centering
\includegraphics[height=190pt]{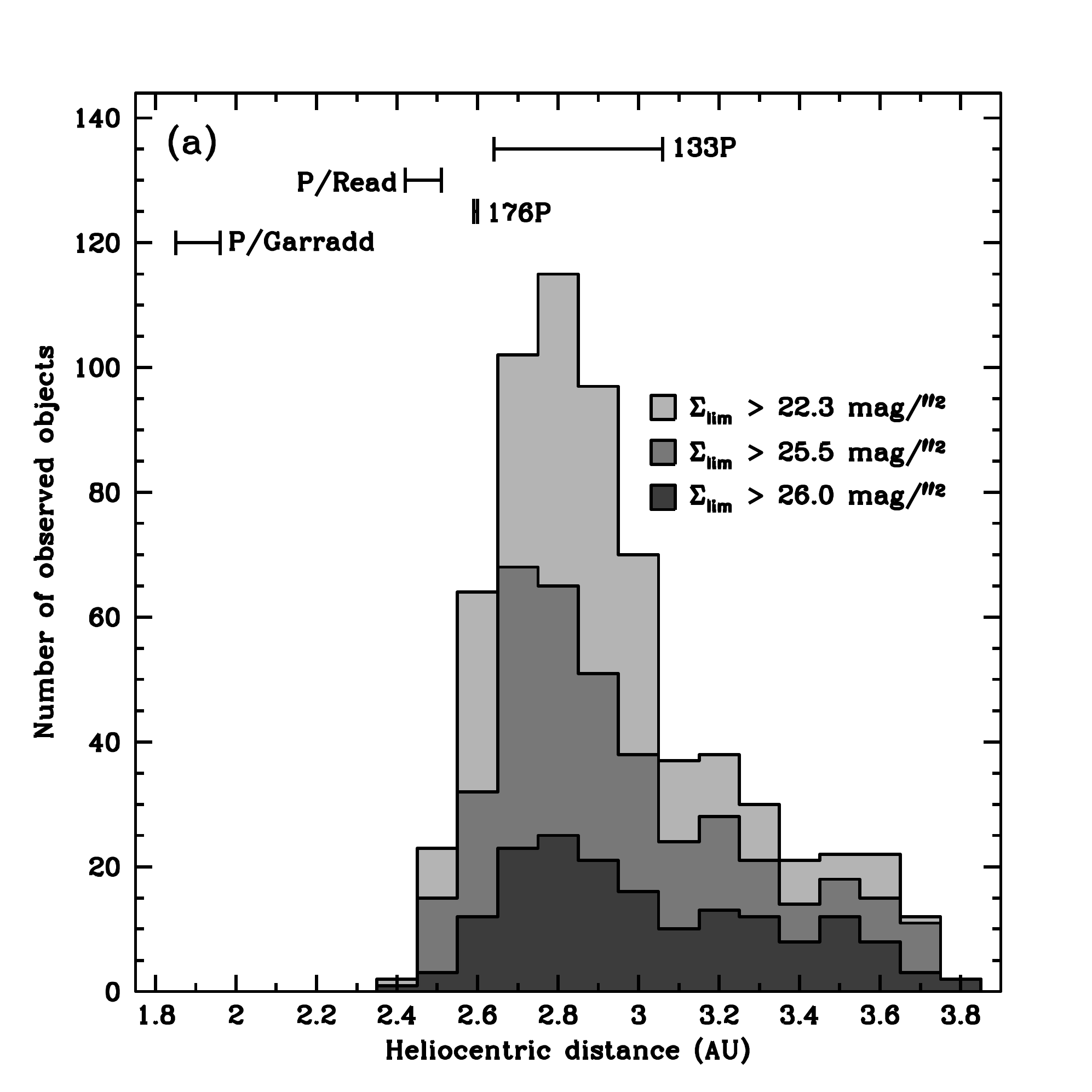}\quad
\includegraphics[height=190pt]{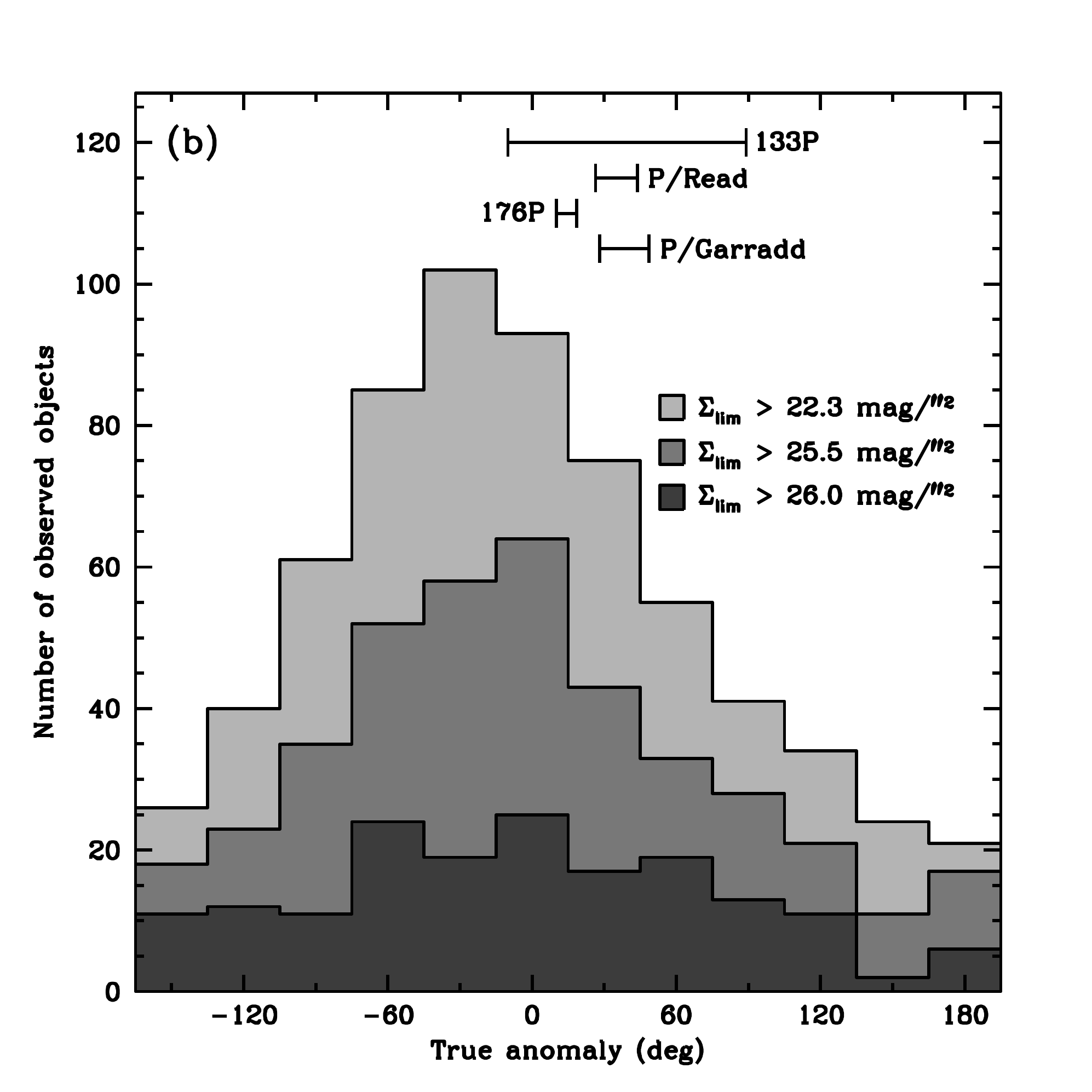}
\includegraphics[height=190pt]{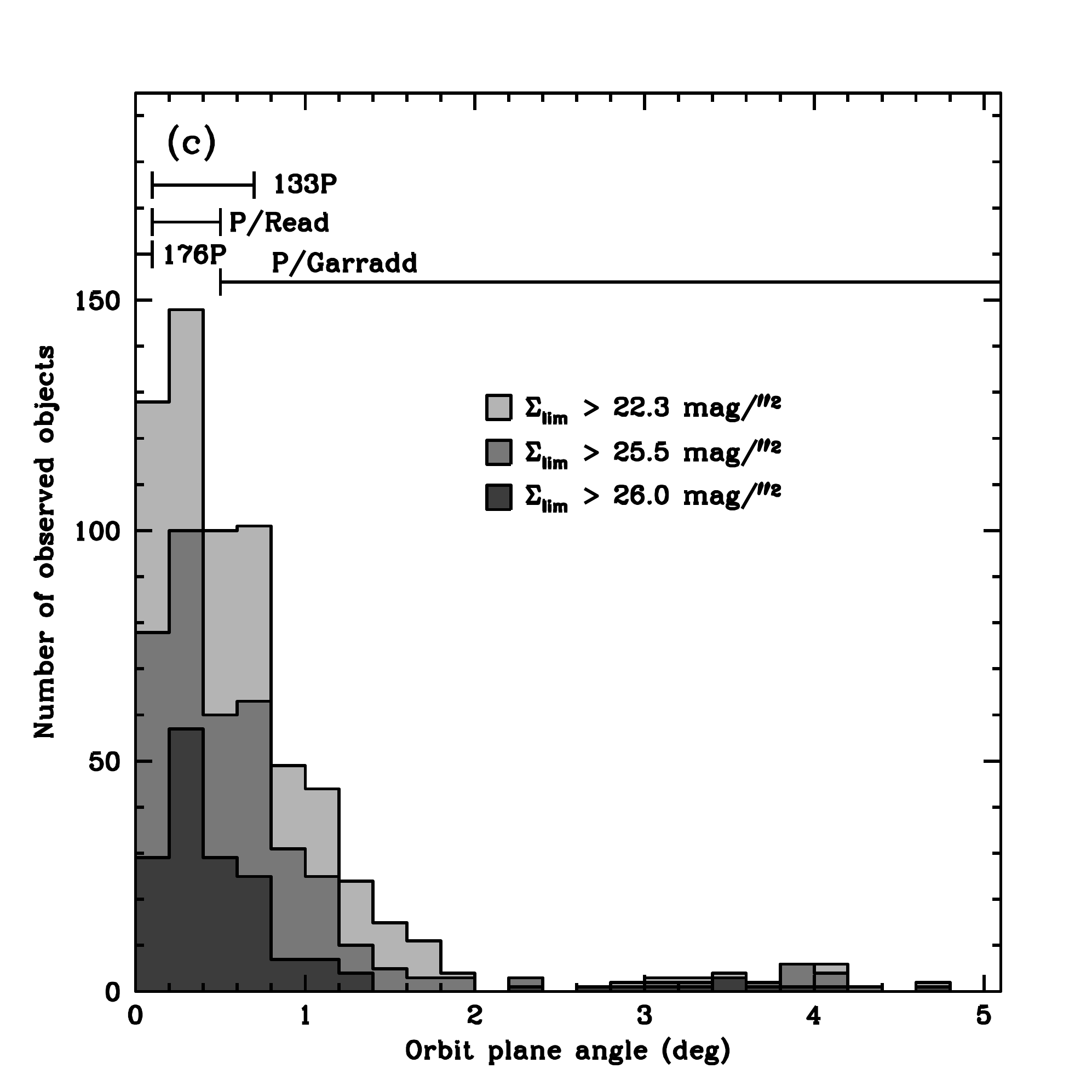}\quad
\includegraphics[height=190pt]{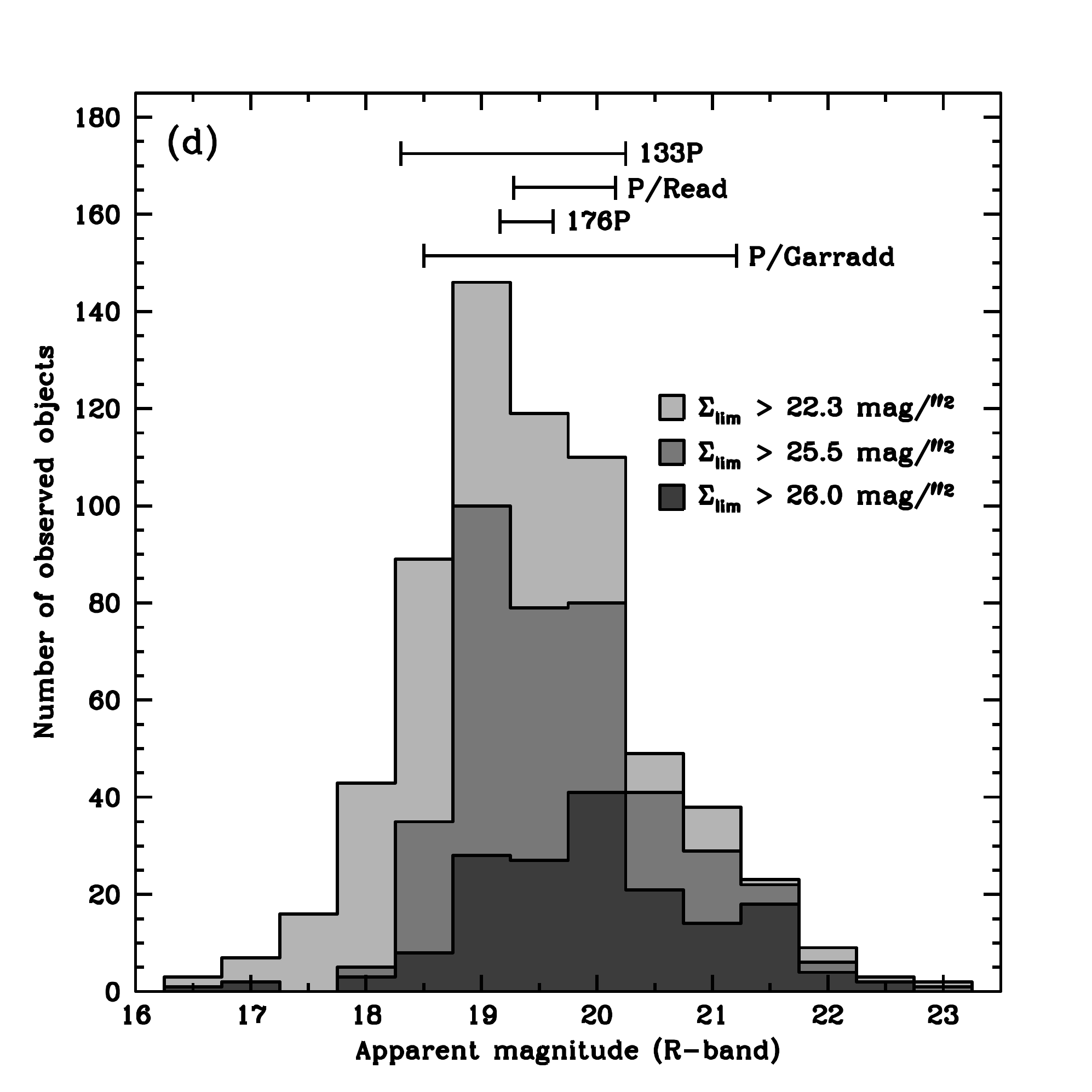}
\caption{Observational circumstances of surveyed targets, including
  (a) heliocentric distance ($R$) in AU, (b) true anomaly ($\nu$) in degrees,
  (c) orbit plane angle ($\alpha_{pl}$) in degrees, and (d) apparent
  $R$-band magnitude.  The observational circumstances of the currently-known
  MBCs during their observed active periods are marked for reference, although
  only 176P was observed as part of this survey.
  Observations for which activity detection limits are fainter than 22.3~mag~arcsec$^{-1}$,
  25.5~mag~arcsec$^{-1}$, and 26.0~mag~arcsec$^{-1}$
  are shown in light gray, medium gray, and dark gray, respectively.
  }
\label{obsproperties}
\end{figure}

Histograms showing the distributions of various physical and dynamical properties and
observational circumstances for surveyed targets are shown in Figs.~\ref{orbphysproperties}
and \ref{obsproperties}.  For reference, the locations of the known MBCs (for
Fig.~\ref{orbphysproperties}) or the range of locations of the known MBCs while
they were observed to be active (for Fig.~\ref{obsproperties}) in these distributions
are marked. These histograms reflect the following intentional selection effects or
observational biases:
\begin{itemize}
\item{The distributions of semimajor axes (Fig.~\ref{orbphysproperties}a),
  eccentricities (Fig.~\ref{orbphysproperties}b), and inclinations
  (Fig.~\ref{orbphysproperties}c) of surveyed targets strongly reflect the
  focus on specific asteroid families, particularly the low-inclination,
  moderate eccentricity Themis family at $\sim3.1-3.2$~AU and the
  low-inclination, low-eccentricity Koronis family at $\sim$2.9~AU
  (also see Figs.~\ref{survey_ae} and \ref{survey_ai}).
}
\item{The distribution of absolute magnitudes (Fig.~\ref{orbphysproperties}d)
  of surveyed targets reflects the focus on asteroids similar to 133P
  \citep[$H_V=15.9$~mag;][]{hsi09a}.  It is also affected, however, by the
  fact that smaller targets are fainter and are more difficult
  to image to the same depth as larger, brighter
  targets.  The known population of small objects is also quite incomplete,
  again due to the faintness of these objects and the difficulty of
  discovering them.
  Larger asteroids were furthermore more likely to have well-known orbits and
  calculated proper orbital elements, necessary for computing family
  linkages on which initial target selection relied.
}
\item{The increased difficulty of observing fainter objects farther from
  perihelion ($\nu=0^{\circ}$) meant that fewer were observed, as can be seen
  from the heliocentric distance and true anomaly distributions of HTP
  observations (Figs.~\ref{obsproperties}a and \ref{obsproperties}b).
  The resulting oversampling of near-perihelion objects
  means that even if the seasonal activity modulation
  hypothesis discussed above (\S\ref{whynoaas}) is correct and there is an
  equal likelihood of activity along an asteroid's entire orbit, active objects
  are still more likely to be near perihelion when they are observed.
  For this reason, the discovery of near-perihelion activity
  in this survey should not be interpreted as confirmation that
  activity in 133P-like asteroids is correlated to perihelion passages.
  Similarly, the correlation of activity of serendipitously discovered MBCs
  to perihelion passages should not necessarily be interpreted as real either
  since this correlation may simply be due to the fact that the objects
  are closer to observers near perihelion and their activity is therefore
  brighter and more easily detected.
}
\item{The low inclinations of most targets
  (Fig.~\ref{orbphysproperties}c) necessarily meant that they
  were also observed at low orbit plane angles (Fig.~\ref{obsproperties}c).
  This trend is significant because low orbit plane angles are where
  dust emission is most easily observed.
  When dispersed by solar radiation pressure, emitted dust
  will tend to spread out in a plane behind an object and aligned
  with its orbit.  Thus, when the Earth is also aligned with the
  object's orbit plane, the optical depth of the emitted dust is maximized.
  Indeed, when observed to be actively emitting dust, the first three known
  MBCs were at low orbit plane angles with respect to the Earth
  ($|\alpha_{pl}|<0.6^{\circ}$).
}
\item{The distribution of apparent magnitudes
  (Fig.~\ref{obsproperties}d) primarily reflects the fact that the
  majority of HTP observations were made using the UH~2.2~m telescope
  and objects with visual $R$-band magnitudes of $18\la m_R\la20$~mag
  were judged to be the most effectively and efficiently observed with
  that telescope aperture.  Fainter asteroids were observed with
  larger telescopes such as Subaru, Gemini, and Keck I, but these targets typically
  were not imaged as deeply as brighter objects due to the desire to also observe as
  many of these smaller asteroids as possible.
}
\end{itemize}

It should be noted that P/Read was serendipitously discovered and so its
strong orbital similarities to 133P and 176P cannot be easily explained
by selection effects or observational biases.
Its discovery may in fact indicate a true predisposition of the
region near the Themis family for harboring
MBCs.  It should also be noted, however, that the discovery of 
P/Garradd in
a completely unexpected, and unsurveyed, region of the main belt underscores the
point that the HTP cannot and should not be viewed as
a representative sampling of the cometary content of the
entire main belt.

Since the discovery of 176P's cometary nature as part of the targeted
HTP, at least two untargeted surveys have been conducted to search for
other MBCs.  \citet{gil09} searched 12390 main-belt
objects observed as part of the Very Wide segment of the Canada-France-Hawaii
Telescope (CFHT) Legacy Survey, finding one known comet (active Centaur 166P/NEAT) and
one unknown comet with an unknown orbit, but no active objects with confirmed
main-belt orbits.  For a small portion of their survey (952 targets),
automated techniques, some similar to those considered for the HTP,
were used to search for cometary activity.  As in the HTP, though,
the majority of their targets (11438 targets) were simply assessed visually.
Both cometary objects were found by visual inspection.
From these results, they estimated an upper limit of
$\sim800$ weakly active MBCs among objects larger than 1.5~km in diameter
with semimajor axes between 3.0~AU and 5.0~AU, consistent with the results found
by the HTP (\S\ref{popsize}).  Meanwhile, \citet{son08} searched 1850 objects observed with the
CFHT as part of the Thousand Asteroid Light Curve Survey \citep{mas09}.
Several different automated techniques were employed to search for activity, but
ultimately no evidence of activity was found in any of those objects.

Such untargeted surveys are clearly better suited
than the HTP for developing generalizations about the prevalence of MBCs
throughout the asteroid belt in that they do not focus on specific regions
of orbital parameter space.
The correlated disadvantage of the untargeted approach, however, that the 
probability of discovering MBCs is much lower than
that of targeted surveys like the HTP, as is reflected by the lack of detections in
either untargeted survey, despite both having much larger target sample
sizes than the HTP.

\subsection{Survey Sensitivity\label{sensitivity}}

The image depth to which candidate objects should be observed
was a key uncertainty at the start of this survey.  When an
active 133P was observed on 2002 November 5, the surface brightness
of a representative section of the trail (just outside the seeing disk
of the nucleus) was $\Sigma=25.2$~mag~arcsec$^{-2}$.
In previous observing runs that fall, 133P's trail was brighter.
Thus, its 2002 November 5 surface brightness could be considered a minimum
surface brightness sensitivity for future observations searching for
similar activity.  This need for deep imaging
had to be balanced with maximizing the target sample size to
maximize the likelihood of finding an object that was actually active.
In the end, $\sim$75\% (488) of the 657 observations made had 3$\sigma$
surface brightness detection limits of $\Sigma_{lim}=25.2$~mag~arcsec$^{-2}$
or better, though many observations went even deeper, with more than
25\% (170 objects) having 3$\sigma$ surface brightness
detection limits of $\Sigma_{lim}>26.0$~mag~arcsec$^{-2}$
(Table~\ref{objprops}).

The discovery of 176P's cometary nature during this survey was made with
two 120~s exposures (which, stacked together, had a 3$\sigma$ surface
brightness detection limit of $\Sigma_{lim}=26.8$~mag~arcsec$^{-2}$)
on 2005 November 26 on the 8~m Gemini telescope.  In these observations,
the tail had a typical surface brightness of $\Sigma=25.3$~mag~arcsec$^{-2}$,
a surface brightness which should have been detectable by $\sim$70\% (466)
of HTP observations.
One cautionary note is that 176P was also observed on
2005 October 24 using the Lulin 1.0~m telescope just one month before the
discovery of its cometary activity by Gemini on 2005 November 26.
No activity was detected in the Lulin data,
which had a surface brightness detection limit of
$\Sigma_{lim}=25.3$~mag~arcsec$^{-2}$, indicating that 176P's tail could
simply have been fainter then, or perhaps poor seeing
($\sim1\farcs5$ for the Lulin data, compared to $\sim0\farcs9$
for the Gemini data) could be to blame for the lack of observed
activity.  Only $\sim$75\% (484) of HTP observations were made on nights
when the seeing was $\leq1\farcs1$, and of these observations, 427
(65\% of the total survey) had $\Sigma_{lim}>25.3$~mag~arcsec$^{-2}$.

To provide a physical basis of comparison, the ratio of dust surface
area to the total nucleus scattering cross-section, $C_d/C_n$, contained within
$1\times10^6$~km at the geocentric distance of each target object,
that is represented by $\Sigma_{lim}$ is also calculated and shown in
Table~\ref{objprops}.  This parameter is roughly correlated with $S/N$ and
is computed to relate the activity detection limit of an observation (which is largely
dependent on the total effective exposure time, size of telescope used, and
proximity of the Moon or bright stars, but not the target itself)
to the size and brightness of the target being observed.
For reference, $C_d/C_n=0.0039$ for 133P's trail when observed by the UH 2.2~m
telescope on 2002 November 5, while $C_d/C_n=0.0022$ for 176P's tail when observed by
Gemini on 2005 November 26.
Thus, activity as intense as 176P's activity, relative to the size and brightness
of the nucleus, should have been detectable in $\sim$70\% of HTP observations
(where 466 of 657 observations had surface brightness detection limits equivalent
to $C_d/C_n\leq0.022$), while activity as intense as 133P's activity, relative to
its nucleus, should have been detectable in $\sim$90\% (588 of 657) of HTP observations.

\subsection{MBC Population Size\label{popsize}}

A natural question to ask following the completion of the HTP
is how many other active MBCs are implied to exist?
This question is rather tricky to answer
given the various ways that HTP results can be scaled to the general
main-belt population and the difficulty of interpreting the statistical
implications of a single detection.  To start out, however,
the probability, $P_{actv}$, that a randomly selected object
in a given population will be active at a given time can be
estimated by
\begin{equation}
P_{actv} = {n_{actv}\over n_{tot}} \cdot f_{orb}
\label{probdet1}
\end{equation}
where $n_{actv}$ is the number of active objects in the population,
$n_{tot}$ is the total number of objects in the population, and
$f_{orb}$ is the fraction of an object's orbit during which it is
actually active.
For this analysis, I adopt the active orbit fraction of 133P of
$f_{orb}\approx 0.25$ \citep[][and references within]{hsi04} to
be typical of all MBCs.

Observationally, $P_{actv}$ can be estimated from
\begin{equation}
P_{actv} = {n_{det}\over n_{surv}}
\label{probdet2}
\end{equation}
where $n_{det}$ is the number of detections of activity in a particular sample set,
and $n_{surv}$ is the total number of candidate objects observed that
belong to that sample set.  Thus, combining Equations~\ref{probdet1}
and \ref{probdet2}, one obtains
\begin{equation}
n_{actv} = {n_{det} n_{tot}\over n_{surv} f_{orb}}
\label{probdet3}
\end{equation}

The HTP detected one active MBC ($n_{det}=1$),
with the survey size ranging from $n_{surv}=657$, if all surveyed objects are
included, to $n_{surv}=427$, if only observations
judged to be sufficiently sensitive and made in sufficiently good seeing conditions to
have permitted the detection of 176P-like activity (\S\ref{sensitivity}) are considered.
Hereafter, only the latter group of objects, where $n_{surv}=427$, is considered.

As of 2009 April 1, there are approximately 450,000 known main-belt
asteroids.  If the entire asteroid belt is considered to be the candidate
population and using $f_{orb}\approx0.25$ (assuming 133P to be typical of all MBCs),
Equation~\ref{probdet3} indicates that there could be $\sim$4200 MBCs in the main
belt.  This result, however, suggests that all main-belt asteroids are 
equally likely to be MBCs, which is likely not the case.

Refining this analysis, one might note that, just like 133P, 176P was
discovered among the highly collisionally-active, C-type-dominated Themis
family, and therefore Themis family targets may be a more appropriate
sample from which to scale these results.  In this case, $n_{tot}=2271$ and
$n_{surv}=262$, giving $n_{actv}\sim35$.
Like scaling to the entire main-belt population, this calculation
assumes that all Themis family members are equally likely to
display activity.  As before, however, this is likely not the case
(\S\ref{whynoaas}): small asteroids are 
more likely to display observable evidence of sublimation due to their
small escape velocities and higher likelihood of being recently
collisionally-produced fragments \citep{che04} that may have near-surface ice deposits.
Thus far, this hypothesis is supported by the evidence.  All four
currently-known MBCs have $H_V>15$.  If only
Themis family asteroids in this size range are considered, using
$n_{surv}(H_V\geq15.0)=205$ and $n_{tot}(H_V\geq15.0)=358$, one finds
that there may only be $\sim$10 MBCs that are detectable via the type of
optical observations that were done in this survey.

Refining this analysis yet again, one might note that unlike 133P and 176P,
P/Read has an osculating eccentricity that places it beyond the
approximate bounds of the Themis family.  It is possible that
when proper elements become available for P/Read, it will
in fact be found to be associated with the Themis family.  Nonetheless,
immediately following P/Read's discovery, I began surveying other
asteroids that, like P/Read, matched the Themis family in semimajor axis and
inclination, but not in eccentricity.  These are the HTP's low-inclination outer-belt
targets.
Using this sample (which only includes objects with
$H_V\geq14.0$) where $n_{tot}=8647$ and $n_{surv}=327$ (the total number
of observed low-inclination targets with $H_V\geq14.0$), I estimate
that $n_{actv}\sim 100$ among kilometer-scale, low-inclination, outer-belt asteroids.
Using analogous lines of reasoning for the other target categories surveyed as part of
the HTP where $n_{det}<1$ and only targets with $H_V\geq14.0$ are considered,
I estimate that $n_{actv}<150$ for the Koronis family, $n_{actv}<40$ for the Karin
cluster, and $n_{actv}<15$ for the Veritas family.

The design of the HTP as a highly biased, targeted survey means that it cannot be used
to infer the prevalence of MBCs in the general main belt population.
There are large regions of the main belt that were not surveyed
({\it cf}., Figs.~\ref{survey_ae} and \ref{survey_ai}) and the recent discovery
of the fourth-known MBC, P/Garradd, shows that MBCs can certainly exist in
those unsurveyed regions (see
Fig.~\ref{orbphysproperties}).  Furthermore, preliminary results suggest that
the nuclei of P/Read and P/Garradd are both extremely small
\cite[sub-kilometer-scale;][]{hsi09b,jew09}, suggesting that, if not for
their prodigious cometary activity, they would never have been discovered
due to their extreme faintness.  These results further suggest
that more as-yet unknown
sub-kilometer-scale cometary asteroids might be discovered
in the future, but without better knowledge of the size of the main-belt
population at this size scale and a systematic, statistically significant
survey of this population, it is difficult
to predict how many of these small MBCs there might be.

\citet{jew09} found that while P/Garradd is characterized by an asteroidal
$T_J=3.217$, its orbit is unstable on a
timescale of $\sim$20~Myr, due largely to the influence of the nearby
8:3 mean-motion resonance (2.706 AU from the Sun) with Jupiter and the $\nu_6$ secular
resonance (near 2.1~AU from the Sun).  Because of these two facts, they hypothesize that the object
may have migrated to its current location from a source region in the
outer asteroid belt, perhaps where the other three known MBCs currently
reside.  This hypothesis is slightly problematic as \citet{dee07} have
actually found that mixing should be minimal among zones of
the main asteroid belt delineated by the strongly chaotic
3:1 (2.5~AU from the Sun), 5:2 (2.8~AU from the Sun), and 2:1 (3.27~AU from the Sun)
mean motion resonances with Jupiter.  In a study of V-type asteroids in
the middle main belt, however, \citet{roi08} found that a non-negligible
number of asteroids could be capable of migrating outward across
Jupiter's 3:1 resonance, and thus it is conceivable that
P/Garradd could be an icy outer-belt asteroid that
analogously migrated inward across Jupiter's 5:2 resonance.

From these results, it is clear that the small number of currently-known
MBCs greatly hinders any attempt to estimate just how common they are
in the main asteroid belt, and that better constraints may not become available
until deep-imaging, all-sky survey telescopes like
Pan-STARRS and LSST discover more.
This discussion has also focused on
currently active MBCs.  If these objects are recently activated
by collisions, there must also be a large population of icy asteroids that have
{\it not} been recently activated.
In the above analysis, all asteroids in a given target category are assumed to be
dormant MBCs, but in truth, the extent of this dormant population is unconstrained,
except that it must assuredly be larger than the active population.
In fact, if recent thermal models are correct, the population of dormant MBCs
could, in principle, span the entire main belt \cite[Fig. 4 in ][]{sch08}.

\subsection{Impact Circumstances and Environment\label{impactenv}}

\citet{hsi04} estimated a peak mass loss rate for 133P in 2002
on the order of $\sim$0.01~kg~s$^{-1}$, corresponding to an
effective area of exposed water ice
of up to a few hundred square meters (assuming a gas-to-dust ratio
of 1), or an active ``vent'' (assumed
here to be circular) with a radius of $r_v\sim10$~m.
Assuming that the vent was collisionally produced, this amount of
exposed ice requires a crater of that size or larger.  The size of
the required crater is difficult to constrain as it depends
on several factors, such as surface topology and the depth of the subsurface ice.
If the layer of inert surface material is thick, volatile
material may only become exposed in the deepest
portion of the crater.
Furthermore, a rapid expected mantling timescale
\citep[$\tau_m\sim15$~days for a $r\sim2$~km nucleus;][]{jew96,jew02}
suggests that much of the ice initially exposed by an activating
impact is likely to be quickly reburied by rubble mantle growth,
necessitating an initial crater that is larger than the calculated
surface area of exposed ice needed to sustain the observed
rate of sublimation.

In order to calculate the size of the impactor required to create
a reasonably-sized vent on an MBC, assuming that these vents are in fact
collisionally-produced, one must first test whether the creation
of 133P's active vent was likely to be gravity-dominated or strength-dominated using
\begin{equation}
R\approx{8G\pi\rho_a^2r_a r_v\over 3Y}
\end{equation}
where $G$ is the gravitational constant,
$\rho_a$ is the asteroid's mean bulk density (taken here to be $\rho_a=1300$~kg~m$^{-3}$),
$r_a$ is the asteroid radius, $Y$ is the strength parameter
(taken here to be $Y\sim5\times10^5$~Pa, a typical value used for porous silicates),
and $R>1$ for gravity-dominated crater
formation and $R<1$ for strength-dominated crater formation \citep{dav99}.
The density used here is a minimum critical density required for 133P to
be stable against centrifugal forces induced by its rapid rotation \citep{hsi04},
but is also consistent with the density of $\rho_a=1300\pm300$~kg~m$^{-3}$
found for (253) Mathilde (another C-type asteroid)
from a flyby by the Near-Earth Asteroid Rendezvous (NEAR) spacecraft
\citep{vev99}.
One finds $R\approx4\times10^{-5}$ for the minimum inferred crater size of $r_v\sim10$~m,
and $R\approx7\times10^{-2}$ for a crater as large as 133P itself
({\it i.e.}, $r_v\sim1.9$~km).  These results indicate that,
due largely to 133P's extremely low surface gravity, any cratering
on the object likely occurs in the strength-dominated regime.

One therefore should use a crater scaling relation for the strength-dominated regime,
\begin{equation}
r_i = {r_v\over C'_D} \left({6\rho_a\over \pi\rho_i}\right)^{1/3} \left({\rho_i v_i^2\over Y}\right)^{\beta/(\beta - 1)}
\end{equation}
where $r_i$ is the impactor radius,
$\rho_i\sim2600$~kg~m$^{-3}$ is the assumed impactor density,
$v_i\sim5$~km~s$^{-1}$ is the assumed impactor velocity,
and $C'_D$ and $\beta$ are material-dependent constants,
adopted here to be $C'_D=1.5$ and $\beta=0.165$ \citep{obr06}.
One thus finds that an impactor with a radius of at least $r_i\sim0.6$~m
is needed to excavate an active site with a radius of $r_v\sim10$~m.
Using this radius as the minimum possible size for the active site,
and taking this calculation as an order-of-magnitude estimate,
hereafter I adopt $r_i\sim 1$~m as the approximate size of
the impactor responsible for collisionally activating 133P, and furthermore,
assume this activating impactor size to also be typical for all other MBCs.

Unfortunately, the population of meter-sized bodies in the asteroid belt
is poorly constrained, meaning that impact rates by such small impactors
are likewise poorly constrained.  Extrapolations of the
known main-belt population (which extends down to approximately kilometer-sized bodies),
however, indicate that
there may be approximately $n_i\sim10^{13}$ meter-sized bodies
in the main belt, where both \citet{che04} and \citet{bot05}
find an approximate sub-kilometer main-belt size distribution function of
$n_i\sim kr_i^{-2.5}$, where $k=3.2\times10^5$.
Approximating the main belt as a cylinder of $r=3.3$~AU and $h\sim1.5$~AU
(where $h$ is the approximate thickness of the asteroid belt measured perpendicular
to the plane of the solar system), with an $r=2.1$~AU cylindrical hole in its center,
and assuming objects are distributed uniformly throughout the main belt,
one derives an approximate impactor number density of $N_i\sim10^{-22}$~m$^{-3}$.

Then, using
\begin{equation}
\tau_{i}\sim(v_{rel}\pi r_a^2N_i)^{-1}
\end{equation}
for 133P with an effective radius of $r_a=1.9$~km and having typical
relative encounter velocities of $v_{rel}\sim5$~km~s$^{-1}$ with other main belt objects
\citep[{\it e.g.},][]{far92a}, one obtains an approximate
impact time scale per object of $\tau_{i}\sim2\times10^{11}$~s, or about one impact
by a meter-sized projectile every 6000~yr ($\sim$1000 orbits).  
This should be regarded as an order-of-magnitude estimate only, as it is based on
an assumption of uniform impactor density throughout the main belt (certainly untrue),
which in turn is based on an extrapolation of the main-belt size-frequency
distribution.
For reference,
following analogous lines of reasoning, order-of-magnitude timescales for impacts
by 0.1-meter-sized and 10-meter-sized projectiles are $\tau_i\sim10$~yr ($\sim2$
orbits) and $\tau_i\sim1$~Myr ($\sim2\times10^5$ orbits), respectively.

\subsection{Active Lifetimes\label{activelifetimes}}

Like typical crater and impactor sizes for the MBCs, the typical lifetime of
an active vent is difficult to constrain and is likely highly variable
for individual situations.
This is in fact the case for all comets \cite[{\it cf}.][]{sek90}, as
numerous variables ({\it e.g.},  location and depth
of the activating crater, local ice content, local surface
topography, orbital obliquity, perihelion distance, nucleus rotation rate)
affect a particular vent's lifetime.  Clearly a vent's active lifetime
depends on the area of ice exposed, the size and morphology
of the underlying volatile reservoir, and the depth of penetration of the
activating impact.
Surface topography near the active site and
orbital parameters will affect the amount of solar radiation received, affecting
the intensity and duration of sublimation activity.  In the case of 133P,
which has a rotation period of 3.471~hr \citep{hsi04}, centrifugal
forces induced by rapid rotation could affect a vent's active lifetime by inhibiting
rubble mantle growth, slowing this means of cometary deactivation.

As directly calculating the lifetimes of active sites on MBCs is difficult to
do with any confidence, an order-of-magnitude estimate can instead be derived
from the statistics of the HTP survey itself.
Earlier (\S\ref{popsize}), it was estimated that 100 of the 8647 kilometer-scale,
low-inclination, outer belt asteroids known at the time when the HTP was in
progress could be currently active.  Making the simplifying assumption
that all asteroids in that population are equally likely to become active
over their lifetimes, one finds that, on average, each individual asteroid is
active over $f_{tot}\sim0.01$ of its total lifetime.  However, 133P is currently
observed to be active over $f_{orb}\sim0.25$ of its orbit.  Thus, assuming
133P to be typical of all kilometer-scale, low-inclination, outer belt asteroids
and if an activating collision occurs every $\sim$1000 orbits, as estimated above,
then the activity initiated by each collision should persist for
$\sim$40 orbits, or about 250~yr.

For MBCs with active orbit fractions of $f_{orb}=0.1$ or $f_{orb}=0.5$, activity
would instead be expected to persist for 600~yr and 120~yr, respectively.
Additionally, if HTP survey statistics are incorrect by a factor of 2 (due, for
example, to the extremely small-number statistics associated with a single detection, 
the uneven sensitivity of HTP observations, or other factors),
the derived active vent lifetime could vary from 60~yr (using $f_{tot}=0.005$ and
$f_{orb}=0.5$) to 1200~yr (using $f_{tot}=0.02$ and $f_{orb}=0.1$).
The typical lifetime of an active vent can therefore be estimated to be
$\tau_{actv}\sim10^2-10^3$~yr.

If it is instead assumed that MBC activity is caused by 0.1-meter-sized
impactors, the higher impact rate of these more populous impactors implies
$\tau_{actv}\sim0.1-1$~yr (following the line of reasoning above, 
using $f_{tot}=0.005-0.02$ and $f_{orb}=0.1-0.5$).
If 10-meter-sized impactors are assumed to be the dominant activating population,
the derived typical active lifetime is $\tau_{actv}\sim10^4-10^5$~yr.  The active lifetime
of 133P has already been observed to be at least 10~yr, well in excess of the
lifetimes implied by 0.1-meter-sized activating impactors.
Meanwhile, the active lifetimes implied by 10-meter-sized activating
impactors are implausibly comparable to or greater than the lifetimes estimated for
classical comets from the outer solar system \citep[$\sim10^4$~yr; {\it e.g.},][]{lev97}
that presumably contain much more ice, spend far smaller fractions of their orbit
actually actively sublimating, and have only recently been perturbed into Sun-approaching
orbits.
Thus, the impactor size implied by the estimated crater size on 133P
(\S\ref{impactenv}) is consistent with impactor sizes implied by HTP statistics.

\subsection{Possible MBC Origin\label{origins}}

Having derived a self-consistent model of the activation and general properties
of MBC activity, however, I turn to the issue of the origin of these bodies
themselves.  The surface area of 133P is approximately given by
$A_S=4\pi r_a^2=4.5\times10^7$~m$^2$.
If a vent with $r_v=10$~m is created every $\sim$6000~yr and
the probability of impacts creating such vents is assumed to be equal over 133P's
surface, one would expect the potentially volatile fraction ($F_{vol}$) of
133P's surface ({\it i.e.}, the portion not yet devolatilized by impacts)
to follow the exponential decay function
\begin{equation}
F_{vol} = e^{-f_At}
\label{devolatization}
\end{equation}
where the fractional devolatilization rate, $f_A$, can be approximated
by $f_A=A_S\tau_i^{-1}(\pi r_v^2)^{-1}$, and $t$ is the elapsed time since 133P's
formation.  One then finds that 133P has a surface devolatization half-life of
$t_{1/2}\sim700$~Myr, and should have less than 10\% of its surface
unaffected by impacts after 2.5~Gyr, its current age if it is assumed to be an
original fragment from the destruction of the Themis family parent body
(\S\ref{surveydesign}).

If 133P is in fact a primordial member of the Themis family, the present-day
timescale for impacts by meter-sized projectiles on still-volatile portions
of its surface should be $\tau_{i,vol} = \tau_i/F_{vol}\sim6\times10^4$~yr.
Assuming the activity detection statistics of the HTP are correct,
this longer impact timescale then implies correspondingly longer
active vent lifetimes
of $\tau_{actv}\sim10^3-10^4$~yr.  Such vent lifetimes are not
inconceivable but may be considered improbable given that they are approaching
the estimated active lifetimes of much icier and better-preserved classical comets
({\it cf}. \S\ref{activelifetimes}).

Thus, for 133P to become activated on a plausible timescale
without requiring implausibly long vent lifetimes,
it may need to be a secondary (or even multi-generation) fragment
from a recent disruption event.  It and other MBCs could be sheared-off surface
fragments or interior fragments of their respective parent bodies.
In either case, the ice that they contain could have previously been located at great
depths in those parent bodies,
preserved against both solar heating and impact devolatilization, and now
only recently found in shallow subsurface reservoirs.

The conclusion that 133P could be a recently-produced fragment of a larger asteroid
is supported by \citet{nes08} who found that 133P may
be a member of the newly-identified Beagle family (named for Themis-family
asteroid (656) Beagle), which they believe formed $<50$~Myr ago from the catastrophic
disruption of a parent object with a diameter of $D>20$~km.  The other MBCs are not
members of this new family nor are they known to be linked with any other known small
families or clusters, but this does not
necessarily preclude their membership in small families that have yet to be
identified.  Studies by \citet{che04} and \citet{bot05} also predict that the
collisional lifetime against destruction of a $D=4$~km asteroid should be
$\sim$2~Gyr, indicating that if 133P and 176P were formed in the initial Themis
family fragmentation event, both should have been collisionally destroyed
by the present day.

\section{Summary}

I describe the design, execution, and analysis of the Hawaii Trails Project
survey that led to the identification of the class of main-belt comets.
I note the following key points regarding the survey itself:
\begin{itemize}
\item{In an effort to find other 133P-like comets in the main asteroid belt,
      657 observations were made of of 599 unique asteroids belonging to five
      major target categories which included the Themis, Koronis, and
      Veritas asteroid families, the Karin cluster (a subset of the
      Koronis family), and kilometer-scale, low-inclination, outer-belt
      asteroids, where the last category also includes all Themis family members
      and was formulated after the discovery
      of the second- and third-known MBCs to match the orbital
      characteristics of all three MBCs known at the time.
}
\item{One surveyed object (176P; also asteroid 118401) was
      found to be active by this survey, and its discovery, in conjunction
      with the serendipitous discoveries of 133P and P/Read,
      led to the identification of the new class of main-belt comets.
}
\item{There are strong observational selection effects, some intentional
      and some unintentional, that must be considered when interpreting
      the statistical implications of the HTP.  Ultimately,
      the HTP should only be considered to be representative
      of the populations of objects actually surveyed.  Even then, it
      should be noted that only one detection was made, and so the
      standard caveats associated with small number statistics apply.
}
\item{The true abundance and distribution of main-belt comets in the
      main asteroid belt may not be truly known until the data from
      deep all-sky survey telescopes like Pan-STARRS and LSST become available.
      These survey telescopes will likely find many more comets in the region
      of the asteroid belt where 133P, P/Read, and 176P are found,
      but may find comets elsewhere as well, such as where P/Garradd is found.
      Any MBCs discovered by these survey telescopes will still
      require targeted follow-up observations by human observers in order to
      confirm the presence of activity and determine physical and photometric
      properties.  Dynamical analysis of newly discovered MBCs will also be
      essential for determining whether they are native to the regions where
      they are discovered or whether they may have migrated from elsewhere.
}
\end{itemize}

I also make the following order-of-magnitude generalizations about the MBCs
from the results of the HTP and currently available data on the currently
best-characterized MBC, 133P (though the
usual caveats associated with small-number statistics apply):
\begin{itemize}
\item{HTP survey statistics imply that there could be $\sim$100 currently
      active MBCs among the kilometer-scale, low-inclination,
      outer belt asteroid population.  
      Additionally, upper limits of 150, 40, and 15 active objects among
      kilometer-scale members of the Koronis family, Karin cluster, and
      Veritas family, respectively, are found.
      The HTP provides no information about the
      prevalence of active MBCs outside these target categories.
      It also provides little information about the size of the population of
      dormant MBCs ({\it i.e.}, asteroids that contain subsurface ice but
      are currently inactive).
}
\item{The estimated size of the active area on 133P is found to correspond
      to an impactor approximately 1~m in size creating an active vent
      approximately 10~m in radius, assuming that the active area
      is collisionally created.  This result is found to be consistent with
      the expected impact rate of projectiles of that size in the main belt
      and the cometary detection rate of the HTP.
}
\item{Assuming that meter-sized impactors are responsible for initiating
      MBC activity, it is estimated that kilometer-scale, low-inclination,
      outer-belt MBCs experience a potentially activating impact
      roughly every 6000~yr and remain active for approximately
      $10^2-10^3$~yr after each collisional activation.
}
\item{The estimated rate of impacts and estimated typical sizes of resulting
      active sites imply that 133P should become significantly
      devolatilized over the course of 2.5~Gyr,
      suggesting that it may not be
      a primordial member of the $\sim2.5$-Gyr-old Themis family.  It is
      suggested instead that 133P, and possibly the other MBCs as well,
      are fragments of more recent disruptions.
}

\end{itemize}

\begin{acknowledgements}
I acknowledge support of this work through a NASA Planetary Astronomy Grant
(via David Jewitt) and STFC fellowship grant ST/F011016/1.
I am grateful to Richard Wainscoat, Rita Mann, Jana Pittichov\'a, and Aaron
Evans for assistance with acquiring observations, Wen-Ping Chen and Wing Ip
at National Central University in Taiwan for access to Lulin Observatory,
Pedro Lacerda, Alan Fitzsimmons, David Asher, and Yan Fern\'andez for valuable
discussion, and an anonymous referee for helpful comments that improved this
manuscript.
I also thank John Dvorak, Ian Renaud-Kim, Dave Brennen, Paul DeGrood, Dan
Birchall, and Ed Sousa at the UH 2.2~m telescope,
Charles Sorensen and Hien Tran at Keck,
Arturo Gomez and Edgardo Cosgrove at Cerro Tololo,
Ming-Hsin Chang at Lulin,
Chad Trujillo, Kathy Roth, and Tony Matulonis at Gemini,
Miki Ishii and Alanna Garay at Subaru, and
Alberto Alvarez, Patricio Ugarte, and the late Hugo Schwarz at SOAR
for technical support of my observations.  
I thank Edward Bowell at Lowell Observatory for making astorb.dat freely
available, and NASA grant NAG5-4741 and the Lowell Observatory endowment
for providing funding for that work.
I also particularly
thank David Jewitt for invaluable advice and support of this work.
\end{acknowledgements}

\clearpage
\newpage

\begin{table}
\begin{minipage}[t]{\columnwidth}
\caption{Telescope Summary}
\label{telescopes}
\centering
\renewcommand{\footnoterule}{}
\begin{tabular}{lllcrr}
\hline\hline
 Telescope
 & Instrument
 & Pix. Scale\footnote{Pixel scale of instrument in arcsec~pixel$^{-1}$}
 & FOV\footnote{Field of view in arcmin in the plane of the sky}
 & Nights
 & $N_{obj}$\footnote{Total number of objects observed} \\
\hline
CTIO 1.0~m   & Apogee512  & $0\farcs47$  & $4\farcm0 \times 4\farcm0$   &  8 &  67 \\
CTIO 1.0~m   & Y4KCam     & $0\farcs289$ & $19\farcm6 \times 19\farcm6$ &  4 &  24 \\
Gemini 8.1~m & GMOS       & $0\farcs146$ & $5\farcm5 \times 5\farcm5$   &  1 &  36 \\
Keck I 10~m  & LRIS (Red) & $0\farcs210$ & $7\farcm2 \times 7\farcm2$   &  2 &  52 \\
Lulin 1.0~m  & VA1300B    & $0\farcs516$ & $11\farcm2 \times 11\farcm2$ &  7 &  36 \\
SOAR 4.1~m   & SOI        & $0\farcs154$ & $5\farcm3 \times 5\farcm3$   &  4 &  28 \\
Subaru 8.2~m & SuprimeCam & $0\farcs20$  & $34'\times27'$               &  1 &  47 \\
UH 2.2~m     & Optic      & $0\farcs14$  & $9\farcm6 \times 9\farcm6$   &  9 &  48 \\
UH 2.2~m     & Tek2048    & $0\farcs219$ & $7\farcm5 \times 7\farcm5$   & 52 & 319 \\
\hline
\end{tabular}
\end{minipage}
\end{table}

\longtab{2}{
\begin{longtable}{llccccr}
\caption{Observing Run Summary\label{obsruns}}\\
\hline\hline
 UT Date
 & Telescope
 & Instrument
 & Weather\footnote{Weather conditions at the time of observations}
 & Moon\footnote{Lunar phase expressed in offset from new Moon
                   (``N'') in days}
 & Seeing\footnote{Typical FWHM seeing in arcsec}
 & $n_{obj}$\footnote{Number of objects observed} \\
\hline
\endfirsthead
\caption{continued.}\\
\hline\hline
 UT Date & Telescope & Instrument & Weather & Moon & Seeing & $n_{obj}$ \\
\hline
\endhead
\hline
\endfoot
2003 Sep 29    & UH 2.2m    & Tek2048    & clear  & N+3   & 1.0 &  3 \\
2003 Sep 30    & UH 2.2m    & Tek2048    & clear  & N+4   & 1.1 &  4 \\
2003 Dec 12    & UH 2.2m    & Tek2048    & clear  & N--11 & 1.1 &  3 \\
2003 Dec 13    & UH 2.2m    & Tek2048    & clear  & N--10 & 1.5 &  1 \\
2004 Apr 14    & UH 2.2m    & OPTIC      & clear  & N--5  & 1.3 &  2 \\
2004 Apr 15    & UH 2.2m    & OPTIC      & clear  & N--4  & 1.5 &  2 \\
2004 Jul 23    & UH 2.2m    & OPTIC      & clear  & N+6   & 0.8 &  1 \\
2004 Jul 24    & UH 2.2m    & OPTIC      & clear  & N+7   & 0.9 &  1 \\
2004 Jul 25    & UH 2.2m    & OPTIC      & clear  & N+8   & 0.6 &  4 \\
2004 Jul 26    & UH 2.2m    & OPTIC      & clear  & N+9   & 0.8 &  2 \\
2004 Aug 08    & UH 2.2m    & OPTIC      & clear  & N--8  & 0.8 & 13 \\
2004 Aug 09    & UH 2.2m    & OPTIC      & clear  & N--7  & 0.6 &  7 \\
2004 Aug 10    & UH 2.2m    & OPTIC      & clear  & N--6  & 0.8 & 16 \\
2004 Aug 13    & UH 2.2m    & Tek2048    & cirrus & N--3  & 1.2 &  4 \\
2004 Oct 14    & UH 2.2m    & Tek2048    & cirrus & N+0   & 1.0 &  8 \\
2004 Nov 04    & UH 2.2m    & Tek2048    & cirrus & N--8  & 1.3 & 10 \\
2004 Nov 05    & UH 2.2m    & Tek2048    & cirrus & N--7  & 1.2 &  6 \\
2004 Nov 06    & UH 2.2m    & Tek2048    & cirrus & N--6  & 1.1 &  6 \\
2004 Nov 07    & UH 2.2m    & Tek2048    & cirrus & N--5  & 1.5 &  3 \\
2005 Apr 10    & UH 2.2m    & Tek2048    & clear  & N+1   & 0.7 &  9 \\
2005 Apr 12    & UH 2.2m    & Tek2048    & clear  & N+3   & 0.9 &  8 \\
2005 Apr 13    & UH 2.2m    & Tek2048    & cirrus & N+4   & 0.8 & 17 \\
2005 Apr 17    & UH 2.2m    & Tek2048    & clear  & N+8   & 1.1 &  2 \\ %
2005 Apr 30-01 & CTIO 1.0m  & 512x512    & clear  & N--8  & 1.4 &  4 \\ %
2005 May 01-02 & CTIO 1.0m  & 512x512    & cirrus & N--7  & 1.2 & 11 \\ %
2005 May 04-05 & CTIO 1.0m  & 512x512    & cirrus & N--4  & 1.9 &  7 \\ %
2005 May 06-07 & CTIO 1.0m  & 512x512    & cirrus & N--2  & 1.7 & 12 \\ %
2005 May 07-08 & CTIO 1.0m  & 512x512    & cirrus & N--1  & 1.9 &  1 \\
2005 May 08-09 & CTIO 1.0m  & 512x512    & cirrus & N+0   & 1.6 & 11 \\ %
2005 May 09-10 & CTIO 1.0m  & 512x512    & cirrus & N+1   & 1.5 &  9 \\ %
2005 May 10-11 & CTIO 1.0m  & 512x512    & cirrus & N+2   & 2.0 & 12 \\
2005 May 16    & UH 2.2m    & Tek2048    & clear  & N+8   & 1.1 &  5 \\
2005 May 27    & UH 2.2m    & Tek2048    & clear  & N--11 & 0.9 & 10 \\
2005 May 28    & UH 2.2m    & Tek2048    & clear  & N--10 & 0.9 & 10 \\ %
2005 May 29    & UH 2.2m    & Tek2048    & cirrus & N--9  & 0.7 &  4 \\ %
2005 May 30    & UH 2.2m    & Tek2048    & cirrus & N--8  & 0.9 & 11 \\
2005 May 31    & UH 2.2m    & Tek2048    & cirrus & N--7  & 0.9 &  5 \\ %
2005 Jun 08    & Keck 10m   & LRIS       & cirrus & N+1   & 1.1 & 19 \\
2005 Jul 06    & UH 2.2m    & Tek2048    & cirrus & N+0   & 1.1 &  7 \\
2005 Jul 07    & UH 2.2m    & Tek2048    & clear  & N+1   & 0.8 &  6 \\
2005 Jul 09    & UH 2.2m    & Tek2048    & clear  & N+3   & 1.7 &  3 \\
2005 Jul 10    & UH 2.2m    & Tek2048    & clear  & N+4   & 0.8 &  7 \\
2005 Aug 25    & UH 2.2m    & Tek2048    & clear  & N--10 & 0.8 & 10 \\ %
2005 Aug 26    & UH 2.2m    & Tek2048    & clear  & N--9  & 0.6 & 13 \\
2005 Aug 27    & UH 2.2m    & Tek2048    & clear  & N--8  & 0.7 &  8 \\
2005 Aug 28    & UH 2.2m    & Tek2048    & clear  & N--7  & 0.9 & 14 \\
2005 Aug 29    & UH 2.2m    & Tek2048    & clear  & N--6  & 0.7 &  8 \\
2005 Aug 30    & UH 2.2m    & Tek2048    & cirrus & N--5  & 0.9 &  6 \\
2005 Sep 01    & UH 2.2m    & Tek2048    & clear  & N--3  & 0.9 &  2 \\
2005 Sep 24    & UH 2.2m    & Tek2048    & clear  & N--9  & 0.9 & 10 \\
2005 Sep 26    & UH 2.2m    & Tek2048    & clear  & N--7  & 1.2 &  5 \\
2005 Sep 27-28 & CTIO 1.0m  & Y4KCam     & cirrus & N--6  & 1.7 &  5 \\
2005 Sep 28-29 & CTIO 1.0m  & Y4KCam     & cirrus & N--5  & 2.5 &  5 \\
2005 Sep 29-30 & CTIO 1.0m  & Y4KCam     & cirrus & N--4  & 2.3 &  7 \\
2005 Sep 30-01 & CTIO 1.0m  & Y4KCam     & cirrus & N--3  & 1.5 &  7 \\
2005 Oct 20    & Lulin 1.0m & VA1300b    & cirrus & N--13 & 1.9 &  5 \\
2005 Oct 22    & Lulin 1.0m & VA1300b    & cirrus & N--11 & 1.7 &  6 \\
2005 Oct 23    & Lulin 1.0m & VA1300b    & cirrus & N--10 & 1.8 &  5 \\
2005 Oct 24    & Lulin 1.0m & VA1300b    & cirrus & N--9  & 1.5 &  8 \\
2005 Oct 25    & Lulin 1.0m & VA1300b    & cirrus & N--8  & 1.5 &  6 \\
2005 Oct 26    & Lulin 1.0m & VA1300b    & cirrus & N--7  & 1.4 &  5 \\
2005 Oct 27    & Lulin 1.0m & VA1300b    & cirrus & N--6  & 1.5 &  1 \\
2005 Nov 02    & UH 2.2m    & Tek2048    & cirrus & N+0   & 1.5 &  2 \\
2005 Nov 03    & UH 2.2m    & Tek2048    & cirrus & N+1   & 1.5 &  6 \\
2005 Nov 04    & UH 2.2m    & Tek2048    & cirrus & N+2   & 1.0 &  6 \\
2005 Nov 05    & UH 2.2m    & Tek2048    & cirrus & N+3   & 0.6 &  6 \\
2005 Nov 06    & UH 2.2m    & Tek2048    & cirrus & N+4   & 0.8 &  9 \\
2005 Nov 07    & UH 2.2m    & Tek2048    & cirrus & N+5   & 0.7 &  4 \\
2005 Nov 21    & UH 2.2m    & Tek2048    & clear  & N--10 & 0.7 &  1 \\
2005 Nov 26    & Gemini 8m  & GMOS       & clear  & N--5  & 0.6 & 36 \\
2005 Dec 24    & UH 2.2m    & Tek2048    & clear  & N--7  & 1.0 & 11 \\
2005 Dec 25    & UH 2.2m    & Tek2048    & clear  & N--6  & 0.9 &  7 \\
2005 Dec 26    & UH 2.2m    & Tek2048    & clear  & N--5  & 1.0 &  8 \\
2005 Dec 27    & UH 2.2m    & Tek2048    & clear  & N--4  & 0.8 &  6 \\
2006 Feb 08    & UH 2.2m    & Tek2048    & clear  & N+10  & 0.9 &  2 \\
2006 Apr 22    & UH 2.2m    & Tek2048    & cirrus & N--6  & 1.1 &  4 \\
2006 Apr 23    & UH 2.2m    & Tek2048    & cirrus & N--5  & 0.9 &  9 \\
2006 Apr 28    & UH 2.2m    & Tek2048    & cirrus & N+0   & 1.0 &  1 \\
2006 Apr 30    & UH 2.2m    & Tek2048    & cirrus & N+2   & 0.7 &  2 \\
2006 May 22    & UH 2.2m    & Tek2048    & cirrus & N--5  & 0.8 &  2 \\
2006 Jun 26    & Subaru     & SuprimeCam & cirrus & N+1   & 0.6 & 47 \\
2006 Jul 01    & UH 2.2m    & Tek2048    & cirrus & N+6   & 0.9 &  3 \\
2006 Oct 17-18 & SOAR       & SOI        & cirrus & N--4  & 1.3 &  2 \\
2006 Oct 18-19 & SOAR       & SOI        & clear  & N--3  & 0.8 &  7 \\
2006 Oct 19-20 & SOAR       & SOI        & clear  & N--2  & 1.1 &  4 \\
2006 Oct 20-21 & SOAR       & SOI        & clear  & N--1  & 0.8 & 15 \\
2006 Nov 11    & UH 2.2m    & Tek2048    & cirrus & N--10 & 1.1 &  2 \\
2007 Jan 27    & Keck       & LRIS       & clear  & N+8   & 0.9 & 33 \\
\hline
\multicolumn{6}{l}{Total number of objects observed}
 ..........................................................................
 & 657 \\
\hline
\multicolumn{7}{l}{$^1$~Weather conditions at the time of observations} \\
\multicolumn{7}{l}{$^2$~Lunar phase expressed in offset from new Moon (``N'') in days} \\
\multicolumn{7}{l}{$^3$~Typical FWHM seeing in arcsec} \\
\multicolumn{7}{l}{$^4$~Number of objects observed} \\
\end{longtable}
}

\begin{landscape}
\begin{table}
\begin{minipage}[t]{\columnwidth}
\scriptsize
\caption{Properties of Objects Imaged}
\label{objprops}
\centering
\renewcommand{\footnoterule}{}
\begin{tabular}{llllccccccclccrrrrrrrrr}
\hline\hline
 Object ID
 & Object
 & $n$\footnote{Number of independent times an object was observed during the course of this survey}
 & Type\footnote{Target category of object (see \S~\ref{surveydesign}) where ``Them'' indicates
		  a Themis family asteroid, ``Koro'' indicates a Koronis family asteroid, ``Veri''
		  indicates a Veritas family asteroid, ``Kari'' indicates a Karin cluster asteroid,
		  and ``Low-i'' indicates a low-inclination outer belt asteroid}
 & Elem.\footnote{Type of orbital elements used, where ``P'' indicates proper elements from the AstDyS
                   website were used, and ``O''
		   indicates osculating elements from JPL's Solar System Dynamics Group's
		   Small-Body Database (http://ssd.jpl.nasa.gov/sbdb.cgi) were used}
 & $a$\footnote{Orbital semimajor axis in AU}
 & $e$\footnote{Orbital eccentricity}
 & $i$\footnote{Orbital inclination in degrees}
 & $T_J$\footnote{Tisserand parameter, computed from listed orbital elements}
 & $H$\footnote{Tisserand parameter, computed from listed orbital elements}
 & UT Date\footnote{UT data in year, month, and day (YYYYMMDD)}
 & Tel.\footnote{Telescope used
     (CT1m: CTIO 1.0~m telescope;
     Gemn: Gemini North 8.1~m telescope;
     Keck: Keck I 10~m telescope;
     Lulin: Lulin 1.0~m telescope;
     SOAR: Southern Astrophysical Research 4.1~m telescope;
     Subr: Subaru 8.2~m telescope;
     UH2.2: University of Hawaii 2.2~m telescope)}
 & $R$\footnote{Median heliocentric distance in AU; from JPL's Solar System Dynamics Group's Horizons ephemeris generator (http://ssd.jpl.nasa.gov/horizons.cgi)}
 & $\Delta$\footnote{Median geocentric distance in AU; from JPL's Horizons ephemeris generator}
 & $\alpha$\footnote{Solar phase angle in degrees; from JPL's Horizons ephemeris generator}
 & $\alpha_{pl}$\footnote{Orbit plane angle (between the observer and object orbit
                   plane as seen from the object) in degrees; from JPL's Horizons ephemeris generator}
 & $\nu$\footnote{True anomaly in degrees; from JPL's Horizons ephemeris generator}
 & $m_R$\footnote{Approximate mean $R$-band magnitude of nucleus as measured
                   from observations}
 & $\sigma_m$\footnote{Estimated magnitude uncertainty, where
		   uncertainties of 0.1~mag are assigned to observations obtained during clear conditions,
                   and uncertainties of 0.5~mag are assigned to observations obtained with cirrus present
		   (observing conditions listed in Table~\ref{obsruns})}
 & $t$\footnote{Total effective exposure time in seconds}
 & $S/N$\footnote{Measured signal-to-noise ratio for composite image}
 & $\Sigma_{lim}$\footnote{Surface brightness detection limit for activity, in mag~arcsec$^{-1}$, as measured
                   in a 1~arcsec$^2$ sampling aperture}
 & $C_d/C_n$\footnote{Fractional scattering surface area of dust with respect to the nucleus cross-section,
                   in $10^{6}$~km$^{-1}$ at the geocentric distance of the nucleus, that is represented by
		   $\Sigma_{lim}$} \\
\hline
20030929\_01 &  48821 (1997 WK$_{35}$)     & 1 & Them  & P & 3.193 & 0.191 &  2.28 & 3.17 & 14.5 & 20030929 & UH2.2 & 2.82 & 2.62 & 20.8 &    0.0 & 299.2 & 19.6 & 0.1 & 1500 &  147 & 25.9 & 0.0009 \\
20030929\_02 &  68899 (2002 JL$_{95}$)     & 1 & Them  & P & 3.091 & 0.146 &  1.48 & 3.21 & 15.3 & 20030929 & UH2.2 & 2.68 & 1.69 &  3.5 &    0.1 &  45.5 & 19.0 & 0.1 & 1200 &  215 & 25.9 & 0.0011 \\
20030929\_03 &  69839 (1998 SJ$_{10}$)     & 1 & Them  & P & 3.047 & 0.157 &  1.81 & 3.22 & 15.2 & 20030929 & UH2.2 & 2.61 & 1.61 &  1.0 &  --0.2 & 317.9 & 18.1 & 0.1 &  900 &  327 & 25.5 & 0.0008 \\
\hline
20030930\_01 &  66065 (1998 RB$_{13}$)     & 1 & Them  & P & 3.115 & 0.147 &  1.47 & 3.20 & 15.2 & 20030930 & UH2.2 & 2.56 & 1.90 & 19.7 &    0.1 & 358.4 & 18.7 & 0.1 &  900 &  155 & 25.7 & 0.0008 \\
20030930\_02 &  69220 (3030 T-3)           & 1 & Them  & P & 3.053 & 0.157 &  1.57 & 3.22 & 15.3 & 20030930 & UH2.2 & 2.50 & 2.06 & 22.9 &    0.2 & 344.1 & 19.4 & 0.1 & 1200 &  106 & 25.3 & 0.0020 \\
20030930\_03 &  71552 (2000 DR$_{7}$)      & 1 & Them  & P & 3.185 & 0.152 &  1.47 & 3.18 & 15.0 & 20030930 & UH2.2 & 2.80 & 1.86 &  8.3 &    0.1 &  57.5 & 18.4 & 0.1 &  900 &  316 & 26.0 & 0.0005 \\
20030930\_04 & 155387 (1994 AC$_{9}$)      & 1 & Them  & P & 3.165 & 0.152 &  1.09 & 3.18 & 16.0 & 20030930 & UH2.2 & 2.63 & 1.69 &  9.7 &    0.9 & 332.7 & 19.3 & 0.1 & 1500 &  140 & 26.0 & 0.0014 \\
\hline
20031212\_01 &  47085 (1999 AW$_{2}$)      & 1 & Them  & P & 3.110 & 0.152 &  1.01 & 3.20 & 14.5 & 20031212 & UH2.2 & 2.68 & 1.71 &  4.8 &  --0.4 &  44.5 & 17.9 & 0.1 &  600 &  206 & 25.3 & 0.0007 \\
20031212\_02 &  68451 (2001 SZ$_{31}$)     & 2 & Them  & P & 3.161 & 0.157 &  0.95 & 3.19 & 14.7 & 20031212 & UH2.2 & 2.71 & 2.79 & 20.6 &    0.2 &   1.2 & 19.6 & 0.1 & 2100 &   83 & 25.9 & 0.0008 \\
20031212\_03 & 139964 (2001 SH$_{7}$)      & 2 & Them  & P & 3.095 & 0.144 &  1.10 & 3.21 & 15.0 & 20031212 & UH2.2 & 2.70 & 2.46 & 21.3 &    0.6 &  18.1 & 20.5 & 0.1 & 2100 &   37 & 25.7 & 0.0025 \\
\hline
20031213\_01 & 112407 (2002 NU$_{39}$)     & 1 & Them  & P & 3.074 & 0.151 &  1.05 & 3.21 & 15.5 & 20031213 & UH2.2 & 2.90 & 1.93 &  3.9 &  --0.4 &  82.5 & 20.1 & 0.1 & 1500 &   63 & 26.1 & 0.0019 \\
\hline
\multicolumn{21}{l}{* Excerpt only;
          full Table~\ref{objprops} available online at the CDS
          (http://cdsweb.u-strasbg.fr/A+A.htx)} \\
\hline
\end{tabular}
\end{minipage}
\end{table}
\end{landscape}

\begin{landscape}
\begin{table}
\begin{minipage}[t]{\columnwidth}
\caption{Survey Summary\label{surveysummary}}
\centering
\renewcommand{\footnoterule}{}
\begin{tabular}{lllrrr}
\hline\hline
  Target Category
  & Definition\footnote{Criteria used to select target list from general
                   asteroid population, where family and cluster members
                   are selected using hierarchical clustering
		   \citep{zap90,zap94} and the maximum allowable
		   $\delta v$ value, $\delta v'$, with respect to each family's or
		   cluster's reference asteroid, is listed}
  & Orbital Element Ranges\footnote{Osculating orbital element ranges for the target samples in this
                   survey, where semimajor axis distances are in AU and
                   inclination values are in degrees}
  & $n_{obj}$\footnote{Number of unique objects observed belonging to each category}
  & $n_{obs}$\footnote{Number of observations of objects belonging to each category (where multiple
                   observations of single targets are counted independently)}
  & $n_{tot}$\footnote{Total number of known potential targets belonging to each category (as of 2009 April 1)} \\
\hline
 Themis family     & $\delta v'=70$~m~s$^{-1}$ with respect to 24 Themis                 & $3.04 < a < 3.27$, $0.05 < e < 0.23$, $0.0 < i <  3.1$ & 342 & 389 &  2271 \\
 Low-$i$ asteroids\footnote{Includes all Themis family members} & $3.0~{\rm AU}<a<3.3~{\rm AU}$, $e<0.3$, $i<3.0^{\circ}$, $H_V>14.0$ & $3.09 < a < 3.21$, $0.03 < e < 0.30$, $0.1 < i <  2.9$ & 426 & 476 & 10929 \\
 Karin cluster     & $\delta v'=10$~m~s$^{-1}$ with respect to 832 Karin                 & $2.86 < a < 2.87$, $0.00 < e < 0.09$, $1.0 < i <  3.2$ &  15 &  15 &   130 \\
 Koronis family\footnote{Includes all Karin cluster members}    & $\delta v'=50$~m~s$^{-1}$ with respect to 158 Koronis               & $2.83 < a < 2.95$, $0.00 < e < 0.12$, $0.7 < i <  3.3$ & 131 & 136 &  2565 \\
 Veritas family    & $\delta v'=40$~m~s$^{-1}$ with respect to 490 Veritas               & $3.15 < a < 3.18$, $0.02 < e < 0.11$, $7.9 < i < 10.6$ &  42 &  45 &   324 \\
 \hline
 Total             &   ... & ...   & 599 & 657 & 13818 \\
\hline
\end{tabular}
\end{minipage}
\end{table}
\end{landscape}

\begin{landscape}
\begin{table}
\begin{minipage}[t]{\columnwidth}
\caption{MBC Properties\label{mbcprops}}
\centering
\renewcommand{\footnoterule}{}
\begin{tabular}{lrrrrrrrrrl}
\hline\hline
 Object\footnote{Proper orbital elements for 133P and 176P from the AstDyS website; osculating
		   elements for P/Read and P/Garradd from JPL's Solar System Dynamics Group's
		   Small-Body Database}
 & $a$\footnote{Semimajor axis in AU}
 & $e$\footnote{Eccentricity}
 & $i$\footnote{Inclination in degrees}
 & $T_J$\footnote{Tisserand invariant, as computed from tabulated orbital elements}
 & $H_R$\footnote{Absolute $R$-band magnitude}
 & $R(active)$\footnote{Range of heliocentric distances, in AU, during which activity has been observed}
 & $\nu(active)$\footnote{Range of true anomalies, in degrees, during which activity has been observed}
 & $\alpha_{pl}(active)$\footnote{Range of orbit plane angles, in degrees, during which activity has been observed}
 & $m_R(active)$\footnote{Range of apparent $R$-band magnitudes when activity has been observed}
 & References\footnote{(1) \citet{els96}, (2) \citet{hsi04}, (3) \citet{hsi09a},
                  (4) \citet{rea05}, (5) \citet{hsi09b}, (6) \citet{hsi06b}, (7) \citet{gar08}, (8) \citet{jew09}} \\
\hline
133P/Elst-Pizarro   & 3.164 & 0.154 &  1.369 & 3.185 & 15.5    & $2.64-3.06$ & $349.9-89.1$ & $0.1-0.7$  & $18.3-20.3$ & 1,2,3 \\
P/2005 U1 (Read)    & 3.165 & 0.253 &  1.267 & 3.153 & 20.1    & $2.42-2.51$ & $26.4-43.9$  & $0.1-0.5$  & $19.3-20.2$ & 4,5 \\
176P/LINEAR         & 3.217 & 0.150 &  1.352 & 3.172 & 15.1    & $2.59-2.60$ & $10.1-18.6$  & $0.0-0.1$  & $19.2-19.6$ & 3,6 \\
P/2008 R1 (Garradd) & 2.726 & 0.342 & 15.903 & 3.217 & $>$17.8 & $1.85-1.96$ & $28.2-48.6$  & $0.5-11.3$ & $18.5-21.2$ & 7,8 \\
\hline
\end{tabular}
\end{minipage}
\end{table}
\end{landscape}

\end{document}